\RequirePackage{fix-cm}
\documentclass[twocolumn]{svjour3}          
\smartqed  
\usepackage{graphicx}
\usepackage{subcaption}
\usepackage{amssymb}
\usepackage{mathtools}
\usepackage{amsfonts}

\usepackage{contour}
\usepackage{xcolor}
\usepackage{blindtext}
\usepackage{afterpage}
\usepackage{boldline}

\usepackage{standalone}
\usepackage{rotating}

\usepackage[ruled,linesnumbered]{algorithm2e}
\usepackage{algpseudocode}
\SetKwComment{Comment}{$\triangleright$\ }{}
\SetSideCommentRight

\newcommand{\opD}{\mathcal{D}}

\newcommand{\opB}{\mathcal{B}}
\newcommand{\opW}{\mathcal{W}}
\newcommand{\opC}{\mathcal{C}}

\newcommand{\opP}{\mathcal{P}}
\newcommand{\opN}{\mathcal{N}}
\newcommand{\opL}{\mathcal{L}}
\newcommand{\bepsilon}{\boldsymbol{\epsilon}}

\newcommand{\Lf}{\pmb{L}}
\newcommand{\Le}{\mathcal{L}}

\newcommand{\wLf}{\contour[1]{black}{$\widehat{L}$}}

\newcommand{\sgn}{\mathfrak{\textbf{sgn}}}
\newcommand{\Vs}{\boldsymbol{\theta}}
\newcommand{\Xy}{\mathfrak{\textbf{z}}}
\newcommand{\xx}{\mathfrak{\textbf{x}}}

\newcommand{\yy}{\mathfrak{\textbf{y}}}

\newcommand{\zz}{\mathfrak{\textbf{z}}}
\newcommand{\bz}{\mathfrak{\textbf{z}}}
\newcommand{\ww}{\mathfrak{\textbf{w}}}
\newcommand{\bw}{\mathfrak{\textbf{w}}}
\newcommand{\bb}{\mathfrak{\textbf{b}}}
\newcommand{\ba}{\mathfrak{\textbf{a}}}
\newcommand{\ccc}{\mathfrak{\textbf{c}}}

\newcommand{\ff}{\mathfrak{\textbf{f}}}

\newcommand{\uu}{\mathfrak{\textbf{u}}}
\newcommand{\bv}{\mathfrak{\textbf{v}}}
\newcommand{\bu}{\mathfrak{\textbf{u}}}
\newcommand{\br}{\mathfrak{\textbf{r}}}
\newcommand{\bd}{\mathfrak{\textbf{d}}}
\newcommand{\bp}{\mathfrak{\textbf{p}}}

\newcommand{\tJ}{\mathfrak{\text{J}}}
\newcommand{\tR}{\mathfrak{\text{R}}}

\newcommand{\tI}{\mathfrak{\text{I}}}

\newcommand{\Dz}{\boldsymbol{\omega}}

\newcommand{\Scale}{\contour[1]{black}{$\zeta$}}

\newcommand{\T}{{\intercal}}
\newcommand{\prox}{\mathfrak{\textbf{prox}}}

\newcommand{\mathR}{\mathbb{R}}

\newcommand{\pel}{\vartheta}

\DeclareMathOperator*{\argmax}{arg\,max}
\DeclareMathOperator*{\argmin}{arg\,min}
\DeclareMathOperator*{\minimize}{minimize}
\DeclareMathOperator*{\subto}{subject\,to}

\newcommand\norm[1]{\left\lVert#1\right\rVert}
\newcommand\abs[1]{\left\lvert#1\right\rvert}
\newcommand{\addpage}{ \newpage\thispagestyle{empty}\mbox{}}

\def\add_pages#1#2{%
  \foreach \index in {1, ..., #1} {%
  \ifthenelse{\boolean{#2}}{\addpage}{} 
  }}

\newcommand{\Stimes}{{\mkern-2mu\times\mkern-2mu}}

\usepackage{array,multirow}

\usepackage[textwidth=20mm]{todonotes}
\let\xtodo\todo
\renewcommand{\todo}[1]{\xtodo[inline,size=\small,color=black!5]{#1}}

\usepackage{environ}
\NewEnviron{gathereq}[1][1]{%
  \begin{equation}
  \scalebox{#1}{$\begin{gathered}\BODY\end{gathered}$}
  \end{equation}
}

\NewEnviron{gather+}[1][1]{%
  \begin{gather}
  \scalebox{#1}{$
  \begin{aligned}[b]
  \BODY\end{aligned}
  $}
  \end{gather}
}

\newboolean{epage}   

\setboolean{epage}{false}

\usepackage[para]{threeparttable}

\usepackage{titlesec}
\titlespacing*{\section}
{0pt}{1.0ex plus 1ex minus .2ex}{1.0ex plus .2ex}
\titlespacing*{\subsection}
{0pt}{1.0ex plus 1ex minus .2ex}{1.0ex plus .2ex}

\usepackage{caption}

\begin{document}
\setlength{\abovedisplayskip}{3pt}
\setlength{\belowdisplayskip}{3pt}

\title{A GPU-Accelerated Light-field Super-resolution Framework Based on Mixed
  Noise Model and Weighted Regularization
}


\author{Trung-Hieu Tran         \and
  Kaicong Sun \and
  Sven Simon
}


\institute{Trung-Hieu Tran \at
  \email{trung-hieu.tran@cis.iti.uni-stuttgart.de}           
}

\date{Received: date / Accepted: date}

\maketitle

\begin{abstract}
  Light-field (LF) super-resolution (SR) plays an essential role in alleviating the current technology challenge in the acquisition of a 4D LF, which assembles both high-density angular and spatial information. Due to the algorithm complexity and data-intensive property of LF images, LFSR demands a significant computational effort and results in a long CPU processing time.
  This paper presents a GPU-accelerated computational framework for reconstructing high resolution (HR) LF images under a mixed Gaussian-Impulse noise condition. The main focus is on developing a high-performance approach considering processing speed and reconstruction quality.
  From a statistical perspective, we derive a joint $\ell^1$-$\ell^2$ data fidelity term for penalizing the HR reconstruction error taking into account the mixed noise situation. For regularization, we employ the weighted non-local total variation approach, which allows us to effectively realize LF image prior through a proper weighting scheme.
  We show that the alternating direction method of multipliers algorithm (ADMM) can be used to simplify the computation complexity and results in a high-performance parallel computation on the GPU Platform.
  An extensive experiment is conducted on both synthetic 4D LF dataset and natural image dataset to validate the proposed SR model's robustness and evaluate the accelerated optimizer's performance.
  The experimental results show that our approach achieves better reconstruction quality under severe mixed-noise conditions as compared to the state-of-the-art approaches.
  In addition, the proposed approach overcomes the limitation of the previous work in handling large-scale SR tasks. While fitting within a single off-the-shelf GPU, the proposed accelerator provides an average speedup of 2.46$\Stimes$ and 1.57$\Stimes$ for $\Stimes 2$ and $\Stimes 3$ SR tasks, respectively. In addition, a speedup of $77\Stimes$ is achieved as compared to CPU execution.
\end{abstract}

\section{Introduction}
\label{intro}
\begin{figure}[t]
  \begin{minipage}[b]{.34\linewidth}
    \centering
    \includegraphics[width=\textwidth]{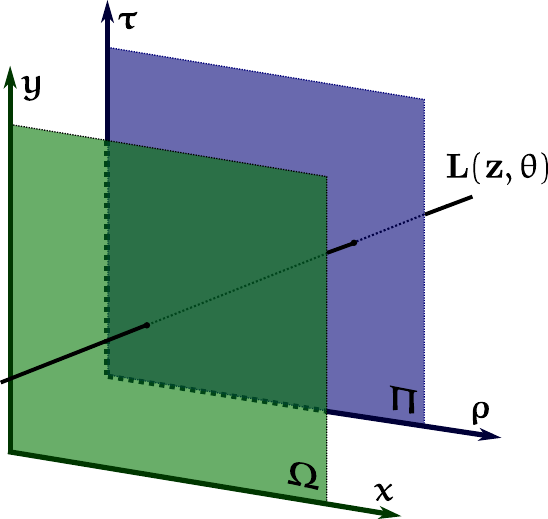}
    \centerline{(a)}\medskip
  \end{minipage}
  \hfill
  \begin{minipage}[b]{0.65\linewidth}
    \centering
    \includegraphics[width=\textwidth]{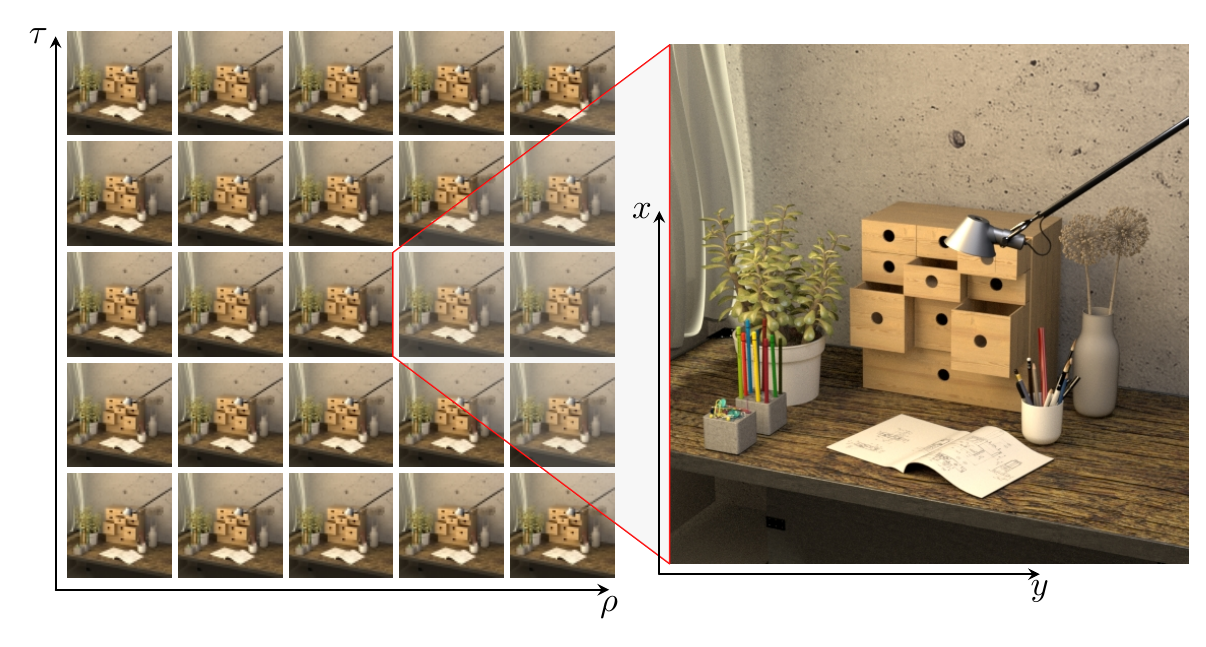}
    \centerline{(b)}\medskip
  \end{minipage}
  \captionsetup{belowskip=-15pt}
  \caption{Light-field representation and acquisition; (a) Two-plane
    parameterisation; (b) 2D array representation of sub-aperture images (SAIs).}
  \label{fig:two_plane}
\end{figure}

Light field (LF) refers to the concept of capturing a comprehensive description of light
rays.
Although a complete parameterization of LF would require a 7D plenoptic function~\cite{Adelson1992single}, practical applications have been successfully made use of its simplified 4D version~\cite{Levoy1996light}.
Among four dimensions, two dimensions are for perspective indexing, and the other dimensions assemble the spatial information, see Fig.~\ref{fig:two_plane}.
The rich-content property of LF brings a great advantage to numerous applications such as in autonomous systems~\cite{Silva2021light}, virtual reality~\cite{Overbeck2018system}, 3D television~\cite{Ni2018360}.
However, this benefit also comes with a cost of computational resources.
Processing 4D LF images requires more memory bandwidth, computing power and runtime than the conventional 2D image. This problem encourages the use of Graphics Processing Unit (GPU) for offloading LF image processing tasks.
There are three main techniques to capture 4D LF data: time-sequential~\cite{Unger2003capturing}, multi-sensors~\cite{Wilburn2005high}, and multiplexing~\cite{Adelson1992single}.
These acquisition methods compromise between spatial resolution and angular or temporal resolution, i.e., using a low-resolution imaging sensor to reduce cost while increasing the number of cameras for a higher angular resolution~\cite{Wilburn2005high}; moving camera with more spatial steps to capture more perspective images but suffering from a long acquisition time~\cite{Unger2003capturing}; Increasing the number of microlenses for a higher spatial resolution while reducing the angular resolution~\cite{Adelson1992single}. These existing challenges in high-resolution LF acquisition are driving recent research on super LF resolution (LFSR)~\cite{Cheng2019light}.

The super-resolution of LF image aims to reconstruct a high-resolution (HR) view, also referred to as sub-aperture image (SAI), from a 2D array of low-resolution (LR) views, see Fig.~\ref{fig:two_plane} (b).
Many approaches have been proposed for the LFSR, including convolutional neural network (CNN) based approaches~\cite{Yuan2018light,Zhang2019residual,Tran20223dvsr} and optimization based approaches~\cite{Bishop2012light,Rossi2018geometry,Alain2018light}.
Although providing high-quality SR results, these approaches typically comprise multiple processing stages and complex algorithms, leading to high computational demand and a long processing time.
For example, multi-stage CNN-based approaches~\cite{Yuan2018light,Tran20223dvsr} divide LFSR into two steps. The first step employs very deep and large CNNs \cite{Kim2016accurate,Lim2017enhanced} for separately up-scaling LR SAIs.
Another refinement CNN is trained and applied in the second step for enhancing the quality of HR SAI.
Other examples are~\cite{Rossi2018geometry} consisting of time-consuming graph processing tasks and~\cite{Alain2018light} involving computational demanding 5D filtering operator.
As far as we know, the literature on GPU accelerated LFSR is very limited despite its importance.
While focusing on the quality aspect, previous approaches put aside the run-time constraint and leave the possibility of accelerating SR tasks undiscussed.

This paper presents a GPU accelerated approach for 4D light-field image super-resolution.
First, we proposed a computational framework for reconstructing high resolution sub-aperture images from 4D LF data, Sec.~\ref{sec:propose}. The LF super-resolution model derived from the statistical perspective consists of a joint $\ell^1$-$\ell^2$ data fidelity term and a weighted nonlocal total variation regularization term. While the first term provides a proper treatment to mixed Gaussian-Impulse noise conditions, the second term introduces an effective way to integrate image features for a better regularization effect. A weighting scheme combining bilateral effect, edge and occlusion features is also proposed.
Secondly, we show that the proposed optimization problem can be effectively solved with the alternating direction method of multipliers (ADMM), Sec.~\ref{sec:sr_opt_ap}. ADMM resolves the main problem of steepest gradient descent in finding a proper step size while avoiding costly line-search operations.
Third, a GPU accelerated architecture is presented for speeding up the iterative solver, Sec.~\ref{sec:sr_gpu_architecture}. Through the realization of transformation matrices with linear functions, which are effectively realized in the form of GPU kernel execution, the proposed approach alleviates the resource shortage of sparse matrix implementation. As shown in the experimental result, the proposed approach can super-resolve large size images (i.e., up to 5760$\Stimes$5760) within a single GPU as compared to 4 GPUs used in the related work~\cite{Sun2021fl}.
In Sec.~\ref{sec:sr_experiment} an extensive experiment is conducted on synthetic 4D LF dataset~\cite{Honauer2016dataset,Shi2019framework} and natural image dataset DIV8K~\cite{Gu2019div8k} to validate the robustness of the proposed SR model and evaluate the performance of the accelerated computational framework. Through the OpenCL framework, the accelerated solver can be deployed on various GPU platforms bringing up a speed-up of 77$\Stimes$ as compared to CPU execution.
The contribution of this work can be summarized as follows:
\begin{itemize}
\item Optimization-based approach for spatially SR of LF image under mixed Gaussian-Impulse noise condition assembling a joint $\ell^1-\ell^2$ data term with weighted nonlocal TV regularization term.
\item Application of ADMM for solving the proposed optimization problem. As shown in Sec.~\ref{sec:sr_opt_ap}, by properly rewriting the optimization problem into the form of ADMM, the solving process is simplified and more suitable for parallel implementation on the GPU platform.
\item OpenCL-based acceleration of the iterative solving process. As discussed in Sec.~\ref{sec:sr_gpu_architecture} and~\ref{sec:sr_experiment}, our accelerator not only provides a significant speed-up as compared to CPU but also overcomes the limitation of the previous work in handling large scale SR problems on the GPU platform.
\end{itemize}
\section{Related Works}
This section discusses the previous works on the super-resolution of 4D LF images, which are divided into two categories: optimization-based approach and learning-based approach. Among these two, learning-based approaches present state-of-the-art performance.
\subsection{Optimization-based Methods}
Optimization-based methods generally formulated LF SR as an optimization
problem, including a data fidelity term built upon a degradation model and a
regularization term based on an assumed prior.
Regarding the data term, previous works proposed either penalizing the coherence
between LR and HR sub-aperture images ~\cite{Bishop2012light,Alain2018light} or
enforcing the intensity similarity over the angular dimension by warping
sub-aperture images~\cite{Tran2018gpu,Rossi2018geometry}.
Regarding the regularization term, the choice is more diverse.
Many image priors are proposed to achieve better output quality and with
reasonable computational cost, i.e.  Markov
Random Field (MRF) \cite{Bishop2012light}, Bilateral TV\cite{Tran2018gpu}, graph-based \cite{Rossi2018geometry},
sparsity \cite{Alain2018light}.

In \cite{Bishop2012light}, Bishop et al. formulated LF imaging process by a set of spatially-variant point
spread functions (PSFs). Under Gaussian optic assumptions, these PSFs are
derived and applied in a Bayesian SR framework.
In \cite{Tran2018gpu}, LFSR was studied in the context of a multi-image
super-resolution problem which considers degradation process as a combination of
three operators: warping, blurring and down-scaling.
The authors employed a variational framework~\cite{Tran2017variational} to
estimate disparity maps  used for warping functions, while BTV
was selected for regularization.
In \cite{Rossi2018geometry}, Rossi et al. assembled an optimization problem with
a graph-based regularizer and  two $\ell^2$ data terms.
They employed block matching for estimating disparity values which was used to
build the graph map.
A patch-based SR approach was proposed in \cite{Alain2018light}.
The authors made used of a 5D transform filter consisting of 2D
shape-adaptive DCT, 2D DCT transform, and 1D haar wavelet.
By a proper selection of 5D patches, a high degree of
sparsity was expected in the transformed signal.
This sparsity property was employed for regularization in combination with a
$\ell^2$ data term.
\subsection{Deep Learning-based Approaches}
Deep learning-based methods for LFSR are mainly categorized
into two groups. While the first group directly exploits the multi-dimensional
structure of LF in learning an end-to-end neural network to synthesize
high-resolution view~\cite{Yeung2018light,Zhang2019residual}, the
second group employs a multi-stages processing model for a step-by-step
improvement of the reconstruction quality~\cite{Yuan2018light,Tran20223dvsr}.
A 4D convolution method was proposed in~\cite{Yeung2018light} to fully exploit
the 4D structure of LF images. The 4D convolution was
realized as an angular-spatial separable convolution allowing the acquisition of
feature maps from both angular and spatial domains.
In~\cite{Zhang2019residual}, a residual CNN-based approach was proposed for
the super-resolution of LF images. Their network was provided with stacking
images from four different angles and predicted an HR image at
the central perspective. Due to the diversity in directional position,
six CNNs were needed for completely reconstructing high-resolution LF.
Compared to learning a single SR network, the two-stages model provides more
flexibility and potentially higher reconstruction quality.
This type of approach takes advantage of well-trained single image super-resolution
(SISR) networks~\cite{Kim2016accurate,Lim2017enhanced} to separately
reconstruct an HR view of each SAI in the first stage.
These HR images are then enhanced in the second stage through a novel
CNN which makes use of inter-perspective information
across multiple SAIs.
In~\cite{Fan2017two}, Fan et al. used VDSR~\cite{Kim2016accurate} in the first
stage and applied a patch-based warping strategy to register the pre-scaling
images. The registered images were combined with a reference image before
feeding to the second-stage CNN for rendering the final HR view.
Yuan et al.~\cite{Yuan2018light} employed EDSR~\cite{Lim2017enhanced} as SISR
and proposed a refinement CNN which relies on 2D epipolar image for the
second stage.
Recently, Tran et al.~\cite{Tran20223dvsr} proposed an approach that exploits
the 3D EPI structure of LF in a two-stages SR framework.
Their method aimed for various LFSR problems, i.e.,
spatial, angular, and angular-spatial super-resolution.
As compared to 2D EPI, which is limited to one spatial dimension, 3D
EPI, which assembles two spatial dimensions along with one angular dimension,
provides a significant contribution to enhance reconstruction quality.
Departed from the usual strategy of employing CNN to directly enhance SR reconstruction quality, Guo et al.~\cite{Guo2021deep} proposed to learn coded aperture from LF data and used it as an implicit LF image prior to a deep learning-based framework for de-noising and reconstructing HR LF. Their approach, however, does not consider Impulse noise and treats de-noising and HR as separate reconstruction problems.
\subsection{GPU Accelerated LF Processing}
The high demand for computational resources due to the large amount of data provided with 4D LF image
encourages the use of GPU as an acceleration platform.
Recent works on GPU-based acceleration focus on two main LF processing tasks, disparity
estimation~\cite{Ivan2018light,Tran2021gvld}, and
super-resolution~\cite{Tran2018gpu}.
For disparity estimation, a GPU
acceleration architecture was presented in \cite{Ivan2018light} for cost-volume based optimization.
The authors employed an advanced matching cost from~\cite{Park2018robust} but decided to
choose the winner-take-all solution over the global minimum as scarification of
accuracy for less complexity and computation.
On the contrary, GVLD~\cite{Tran2021gvld} proposed a GPU-accelerated approach based on
a variational computation framework. The framework combines the intrinsic
sub-pixel precision of variational formulation and the effectiveness of weighted median filtering to
produce a highly accurate solution.
A fully parallelized and optimized OpenCL implementation was provided for
finding the global minimum solution.

For super-resolution, Tran et al.~\cite{Tran2018gpu} proposed to accelerate the
optimization problem which assembles an $\ell^1$ data fidelity term and
a BTV~\cite{Farsiu2004fast} regularization term.
Using steepest descent as an iterative solver, which is fully realized with
OpenCL kernel execution, the proposed approach provides a significant
speed-up as compared to the implementation running on CPU.
This paper extends our previous work~\cite{Tran2018gpu} mainly as follows.
First, we revisit the super-resolution model from the statistical
perspective and propose a mixed noise (Gaussian and Impulse noise) model
based on a combination of $\ell^1$ and $\ell^2$ fidelity terms.
Secondly, we propose a nonlocal total variation weighting scheme that combines
bilateral filtering with image features to improve the regularization effect.
Thirdly, the alternating direction method of multipliers (ADMM) is employed in this work for
solving the optimization problem as a replacement to the steepest descent.
ADMM address the short-coming of the steepest descent in finding appropriate
step-size while avoiding time-consuming line-search.
Lastly, we present an accelerated architecture for realizing the computational
framework on the GPU platform.
The proposed approach is validated and evaluated through an extensive
experiment on synthetic 4D LF dataset and high-resolution natural image dataset.
\section{Proposed Approach}
\label{sec:propose}
This section discusses our proposed approach for reconstructing high-resolution LF images under mixed noise conditions. The section starts with a presentation of the degradation model and notation, which form a basis for discussing the proposed optimization model derived from the Bayesian image reconstruction framework. Our selection of data fidelity term and regularization term are consecutively discussed at the end of this section.
\subsection{Degradation Model and Notation}
\label{sec:sr_degrad}
Light-field is a 4D parameterization of the plenoptic
function~\cite{Levoy1996light} which can be illustrated as a light ray intersecting with two parallel
planes,
\begin{equation}
  \Lf\colon \Omega \Stimes \Pi \rightarrow \mathbb{R},\qquad  (\Xy,\Vs) \rightarrow
  \Lf(\Xy,\Vs),
\end{equation}
with $\Vs = [\rho,\tau]^T$ and $\Xy =[x,y]^T$ indicate the coordinates in the directional plane $\Pi \subset
\mathbb{R}^2$ and the spatial plane $\Omega \subset \mathbb{R}^2$,
see Fig.~\ref{fig:two_plane}(a).
By fixing the directional coordinate $\Vs$ and let spatial coordinate $\Xy$
vary, we obtains the spatial information from one perspective.
Such spatial information is referred to as a sub-aperture image (SAI) or a
perspective image.
Fig.~\ref{fig:two_plane}(b)
shows a $5\Stimes 5$ angular views of LF scene `\textit{table}'~\cite{Honauer2016dataset}.
From this perspective, a 4D LF is a collection of 2D images captured from
different viewpoints and the reconstruction of high-resolution SAIs shows a
strong connection to the multi-image super-resolution (MISR) problem.
\begin{figure}
  \centering
  \includegraphics[width=\linewidth]{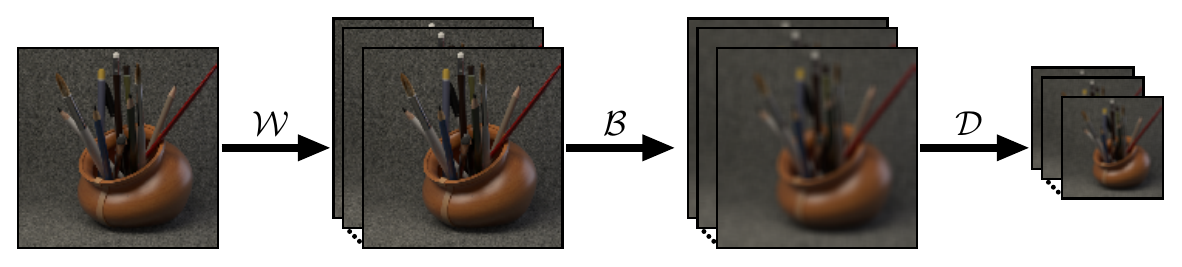}
  \captionsetup{belowskip=-15pt}
  \caption[]{Degradation process}
  \label{fig:sr_degradation}
\end{figure}

Let us rearrange the 2D angular view of LR SAIs into a 1D set of
$s_k$ LR observations $Y_k \in \mathbb{R}^{s_y\Stimes s_x}$, $k \in
[1,s_k]$.
Our goal is to approximate the HR version
$X \in \mathbb{R}^{s_Y\times s_X}$, where $s_y\times s_x$ and $s_Y\times s_X$
are the size of LR images and the size of HR image,
respectively.
In practice, a LR image $Y_k$ is considered as a
degraded version of the HR image $X$. This degradation can be
modelled by the application of three linear operators: warping ($\opW_k$),
blurring ($\opB$), and down-sampling ($\opD$), as depicted in
Fig.~\ref{fig:sr_degradation}. The warping operator represents the positioning
of the camera. Shifting the camera's position will result in the corresponding
shifts of pixels in the captured image.
We define the warping operator as $\opW_k: \mathR^{s_Y\times s_X} \rightarrow
\mathR^{s_Y\times s_X}$ which transforms a HR image into a new one
observed from a different perspective. The blurring operation represents the
point spread function (PSF) which describes the response of an imaging system.
Depending on the setup of lenses and imaging sensors, PSFs can
be very complicated and even spatially variant. However, as shown in the
literature~\cite{Farsiu2004fast,Sun2020multi},
it is sufficient to assume a spatially invariant version of PSF which
can be modelled by a linear operator, i.e. $\opB: \mathR^{s_Y\times s_X}
\rightarrow \mathR^{s_Y\times s_X}$. The down-sampling operator represents the
digital sampling process of an imaging sensor, i.e., $\opD: \mathR^{s_Y\times s_X}
\rightarrow \mathR^{s_y\times s_x}$.
As a combination of these linear operators, the image foundation process can be described as
\begin{equation}
  \label{eq:sr_model_op}
  Y_k =  \opD_k\circ \opB \circ \opW_k (X)  + \epsilon_k, \forall k \in [1,s_k],
\end{equation}
where $\epsilon_k$ represents the measurement error or the additive noise which is
practically assumed to follow Gaussian distribution or Laplace distribution.
For a better presentation, we transform Eq.~\ref{eq:sr_model_op} into vector form,
\begin{equation}
  \label{eq:sr_model_vec}
  \yy_{k} = DBW_k\xx + \bepsilon_k
\end{equation}
where $\yy_k, \bepsilon_k \in \mathbb{R}^{s_xs_y}$ and $\xx \in
\mathbb{R}^{s_X s_Y}$ are the column-vector representations of $Y_k,
\epsilon_k$ and $X$.
Linear transformation matrices $D$, $B$, and $W_k$ respectively replaced the linear operators
$\opD$, $\opB$, and $\opW_k$.
To further simplify the notation, we define $p=s_Xs_Y$, $q=s_xs_y$, and combine $D,B$
and $W_k$ into $A_k$, i.e., $A_k=DBW_k$.
It follows that  $B,W_k \in \mathR^{p\times p}$, $D \in
\mathR^{q\times p}$, and $ A_k \in \mathR^{q\times p}$.
\subsection{Bayesian Image Reconstruction Framework}
Let us start with the standard Bayesian formulation
which poses the SR problem as a maximum a posteriori (MAP) estimation of
HR image $\xx$ given a set of LR samples $\{\yy_k\ |\
k=1,..,s_k\}$:
\begin{equation}
  \tilde{\xx} = \argmax\limits_{\xx} \opP(\xx|\yy_i,...,\yy_{s_k}),
\end{equation}
where $\opP(\xx|\yy_i,...,\yy_{s_k})$ is called posterior and represents the
conditional probability density of $\xx$ given the set of degraded images ($\yy_k$).
Follow Bayes' rule, we have
\begin{equation}
  \opP(\xx|\yy_1,...,\yy_{s_k}) = \frac{\opP(\xx)\prod\limits_{k=1}^{s_k} \opP(\yy_k|\xx)}{\prod\limits_{k=1}^{s_k}\opP(\yy_k)},
\end{equation}
with $\opP(\yy_k|\xx)$ is a likelihood function which encodes the likelihood that the
HR image $\xx$ is due to the LR observation $\yy_k$.
This function is defined based on the assumption of the noise model of
$\bepsilon_k$. Here, we assume that the noise
effecting the observed LR image $\yy_k$ is independent.
$\opP(\xx)$ is an image prior describing the properties of the high-resolution
image being reconstructed.
Since the low-resolution samples are known, $\opP(\yy_k), k=1,...,s_k$, are
constants, and the above MAP problem can be transformed into a minimization of negative
log-likelihood
\begin{gather+}[0.9]
  \argmax\limits_{\xx} \opP(\xx|\yy_1,...,\yy_{s_k}) = \argmin\limits_{\xx}
  - ln \opP(\xx)  - \sum_{k=1}^{s_k}ln \opP(\yy_k|\xx).
  \nonumber
\end{gather+}

The above two logarithmic terms represent the typical setup of an optimization problem
consisting of a data fidelity term (i.e., $E(\xx) := - \sum_{k=1}^{s_k}ln \opP(\yy_k|\xx)$) and a
regularization term (i.e., $ R(\xx) := - ln \opP(\xx)$)
\begin{equation}
  \begin{aligned}
    \hat{\xx} =\quad \argmin\limits_{\xx} E(\xx) + R(\xx)
  \end{aligned}
  \label{eq:sr_d_r}
\end{equation}
\subsection{The Data Fidelity Term}
The construction of the data fidelity term depends on the noise models which are
practically assumed to follow Gaussian and Laplace distribution~\cite{Rodriguez2013total}.
For additive Gaussian noise, $\bepsilon_k \sim \opN(\mu_k ,\sigma_k^{2})$
follows a normal distribution with the probability density function given by
$\left(1/\sqrt{2\pi\sigma_k^2}\right) e^{-(\mu_k - \bepsilon_k)^2/2\sigma_k^2}$. Assuming a
$0$ central distribution (i.e. $\mu_k=0$), the likelihood function
$\opP(\yy_k|\xx)$ reads
\begin{equation}
  \opP(\yy_k|\xx) \propto \texttt{exp}\left(\sum_{k=1}^{s_k}\norm{A_k\xx - \yy_k}_2^2\right),
\end{equation}
which results in a well-known least square fidelity term.
In the case of Laplace noise (i.e., impulse noise), $\bepsilon_k\sim \opL(\mu_k,
b)$ has the probability density function given by $\frac{1}{2b} e^{-\norm{\mu_k
    - \bepsilon_k }_1/b} $,
\begin{equation}
  \opP(\yy_k|\xx) \propto \texttt{exp}\left(\sum_{k=1}^{s_k}\norm{A_k\xx - \yy_k}_1\right).
\end{equation}
This results in an $\ell^1$ norm data fidelity term, which shows robustness against outliers and superior
performance with impulse noise \cite{Farsiu2004fast}.
In order to handle the mixed Gaussian-impulse noise situation, we followed the
previous works~\cite{Jia2016image,Hakim2020multi} to combine $\ell^1$ and $\ell^2$ norm
resulting in a joint $\ell^1-\ell^2$ data fidelity term,
\begin{equation}
  E(\xx) = \sum_{l\in\{1,2\}} \lambda_l\sum_{k=1}^{s_k}\norm{A_k\xx - \yy_k}_l^l,
\end{equation}
with parameters $\lambda_1$ and $\lambda_2$ control the contribution of $\ell^1$
and $\ell^2$ norm respectively.
\subsection{Regularization Term}
\label{sec:sr_opt_reg_term}
In Bayersian framework, it is generally assumed
that $\xx$ is an Markov random field (MRF) with a strictly positive joint
probability density.
Therfore, following Hammersly-Clifford theorem, its joint probability density
must have the form of a Gibbs
distribution\cite{Geman1984stochastic}:
\begin{equation}
  \label{eq:sr_gibbs}
  \opP(\xx) \propto \frac{1}{Z} \texttt{exp}\left( -\frac{1}{T} \sum_{C\in \opC}V_{C}(\xx) \right),
\end{equation}
where $Z$ is a normalizing constant, $T$ stands for \textit{temperature}
and controls the degree of peaking~\cite{Geman1984stochastic}.
$V_{C}$ is called potential defined for a local group of pixels or clique $C$.
The sum is for a set $\opC$ of all possible cliques.
The definition of clique set $\opC$ and the selection of the potential $V_{C}$ lead to
various types of image prior, which share the following form
\begin{equation}
  \label{eq:sr_gibbs_gen}
  \opP(\xx) \propto \texttt{exp}\Big( \sum\limits_{\uu\in\Omega} \sum\limits_{\bv \in \opN(\bu)} w(\bu,\bv) \Phi(\xx_{\bu}, \xx_{\bv}) \Big),
\end{equation}
where $\bu,\bv \in \Omega$ represents the 2D
indices of $\xx$.
$w:\Omega\Stimes \Omega \rightarrow \mathbb{R}^+$ and $\phi: \mathbb{R}\Stimes\mathbb{R} \rightarrow
\mathbb{R}^+$ are respectively weighting function and distance function.
The weighting function characterizes the dependency in pixel locations, while the
distance function penalizes the difference in pixel intensities.
$\opN(\bu)$ represents a set of indices defined with regarding to the index
$\bu$.
By setting $\Phi$ to the absolute difference, we come to the following
weighted regularization term,
\begin{equation}
  \label{eq:sr_reg_final}
  \tR(\xx) = \sum_{\bu\in\Omega} \sum_{\bv \in \opN(\bu)} w(\bu,\bv) |\xx_{\bu} - \xx_{\bv}|,
\end{equation}
which can be considered as a generalized version of many total variation based
image priors, i.e. TV~\cite{Rudin1992nonlinear}, BTV~\cite{Farsiu2004fast}, NLTV~\cite{Gilboa2008nonlocal},
and BSWTV~\cite{Sun2021bilateral}.
In the vector form, Eq.~\ref{eq:sr_reg_final} can be rewritten as
\begin{equation}
  \tR(\xx) = \sum_{\bv \in \opN(\bu)} \norm{W_{\bd} \odot(S_{\bd} - \text{I})\xx}_1,\quad \bd=\bu-\bv,
  \label{eq:sr_reg}
\end{equation}
where $S_{\bd}\in \mathbb{R}^{p\times p}$ denotes the shifting matrix which
shift $\xx$ by $\bd$ (in 2D coordinate), $\odot$ denotes the Hadamard product.
Weighting functions are assembled in weighting matrix $W_{\bd}=
\text{diag}(\bw_{\bd})$, with $\bw_{\bd}\in \mathbb{R}^{p}$.
The main advantage of this regularization term is the flexibility in defining
weighting function to capture unique feature of the SR problem.
For example, setting $\opN(\bu)$ to direct neighborhood and
the weighting to a constant gives us TV~\cite{Rudin1992nonlinear} which regularizes the local smoothness between
adjacent pixels. Setting weighting to a function of the pixel distance give us
BTV~\cite{Farsiu2004fast}, which assumes that the smoothness is spatially dependent.
Another weighting scheme based on bilateral spectrum used in
~\cite{Sun2021bilateral} provides a successful regularization for mixed
Gaussian-Poisson noise images.
Considering the 4D LF data, we proposed a discontinue-aware weighting
scheme which assemble three data properties, i.e., spatial distance, edge and  occlusion feature,
\begin{equation}
  \bw_{\bd} := w_d \bw_{e}\odot \bw_{o},\quad w_{\bd}\in \mathbb{R},\bw_{e},\bw_{o}\in\mathbb{R}^p,
\end{equation}
where the spatial weight $w_{d} := \texttt{exp}\left( \frac{\norm{\bd}_2^2}{\sigma_s}
\right)$ adjusts the impact of weighting w.r.t. the relative distance $\bd$ and
provide a bilateral filtering effect.
The edge weight $\bw_{e}:= \texttt{exp}\left( \frac{\norm{\nabla
      \xx}_2^2}{\sigma_e} \right)$ and the occlusion weight $\bw_{o}$ penalize
the smoothness at image discontinuing area.
We follow the related works~\cite{Sand2008particle,Tran2021gvld} to define the
occlusion weight $\bw_{o}$ as follow,
\begin{equation}
  \label{eq:weight_occ}
  \bw_o(\Xy) = e^{-\frac{b(\Xy)^2}{2\sigma_{o_1}^2}}
  e^{-\frac{p(\Xy)^2}{2\sigma_{o_2}^2}},
\end{equation}
where $b(\Xy)$ and $p(\Xy)$ are the functions of occlusion boundary and
projection error respectively. By an one-side divergence, $b(\Xy)$ provides
weighting to occluding boundary,
\begin{equation}
  b(\Xy) = \left\{
    \begin{array}{ll}
      \texttt{sum}\{\nabla\Dz(\Xy)\} , & \texttt{sum}\{\nabla\Dz(\Xy)\}<0\\
      0 , & \text{otherwise}\\
    \end{array}
  \right.,
\end{equation}
where $\nabla\Dz$ denotes the gradient of a disparity map.
The projection error function is computed as the intensity difference between
a warped view and the reference view, i.e., $p(\Xy) = \Lf(\Xy,\Vs_0) - \Lf(\Xy +\Vs_i\Dz(\Xy),\Vs_i)$
\section{Optimization Approach}
\label{sec:sr_opt_ap}

Combining the data-fidelity term and regularization term discussed in the
previous section, we finalize the minimization problem with the following cost
function
\begin{equation}
  \begin{split}
    \tJ(\xx) =&\lambda_1\sum\limits_{k=1}^{s_k}\norm{A_k\xx - \yy_k}_{1} +
    \lambda_2\sum\limits_{k=1}^{s_k}\norm{A_k\xx - \yy_k}_{2}^{2} \\
    &\quad+ \sum\limits_{d=1}^{s_d} \norm{W_d\odot(S_{d} - \tI)\xx}_1,
  \end{split}
  \label{eq:sr_fin}
\end{equation}

Although non-smooth, the cost function is convex, and the existence of the
global minimized solution is guaranteed. There are many algorithms that can be
used to optimize it. One of the traditional approaches to solving this problem
is applying a first-order iterative algorithm such as steepest gradient descent.
A more recent approach is alternating direction method of multipliers
(ADMM)~\cite{Boyd2011distributed}, which breaks a complex optimization problem
into smaller sub-problems, each can be solved in a simpler manner. Although ADMM
requires more computation for each iterative step as compared to gradient
descent, we notice that the overall computation of ADMM is much less considering
the similar minimization threshold.
We start with rewriting the objective function into a more compact form,
\begin{equation}
  \tJ(\xx)= \norm{A\xx-\bb}_2^2 + \norm{F\xx - \bb'}_1,
  \label{eq:sr_l1l2l1_obj}
\end{equation}
the matrices $F$, $A$ and columns vectors $\bb$, $\bb'$ are defined as in Eq.~\ref{eq:sr_admm_compact}.
Notice that $\lambda_1$ and $\lambda_2$ are absorbed into the matrices and column vectors for simplifying the notation.
The sizes of $A$, $F$, $\bb$ and $\bb'$ are respectively $qs_k\times p$, $(qs_k+ps_d)\times p$, $qs_k\times 1$ and $(qs_k+ps_d)\times 1$.
All transformation matrices ($A_k$ and $S_d$) and  weighting matrices ($W_d$) are assembled into $A$ and $F$. Low-resolution images $b_k$ are stacked into $\bb$.
$O_{ps_d}$ is zero vector with the size of $ps_d\times 1$.
\begin{gather+}[0.88]
  &A := \sqrt{\lambda_2} \begin{bmatrix}
    A_1\\ A_2\\ ... \\ A_{s_k}
  \end{bmatrix},\quad
  \bb := \sqrt{\lambda_2}\begin{bmatrix}
    \yy_1 \\ \yy_2 \\ ... \\ \yy_{s_k}
  \end{bmatrix}\\
  &F := \begin{bmatrix}
    \frac{\lambda_1}{\sqrt{\lambda_2}} A \\ S
  \end{bmatrix},
  \bb' := \begin{bmatrix}
    \frac{\lambda_1}{\sqrt{\lambda_2}}\bb \\ O_{ps_d}
  \end{bmatrix},
  S := \begin{bmatrix}
    W_1\odot (S_1-I) \\
    W_2\odot (S_2-I)\\
    ... \\
    W_{s_d}\odot (S_{s_d}-I)
  \end{bmatrix}
  \label{eq:sr_admm_compact}
\end{gather+}

Taking the compact representation, we rewrite the optimization problem in
Eq.~\ref{eq:sr_fin} into the form of ADMM problem,
\begin{equation}
  \begin{split}
    &\minimize_{\xx,\zz} \quad \norm{A\xx - \bb}_2^2 + \norm{\zz}_1\\
    &\subto \quad F\xx - \zz = \bb',
  \end{split}
  \label{eq:sr_admm_l1l2l1_prob}
\end{equation}
with the augmented Lagragian reads,
\begin{gather+}[0.95]
  \Le_{\pel}\big(\xx, \zz, \ww\big)& :=\norm{A\xx - \bb}_2^2  + \norm{\zz}_1 \\
  &+ \ww^\T(F\xx - \zz -\bb')
  + \frac{\pel}{2} \norm{F\xx - \zz -\bb'}_2^2
\end{gather+}

The ADMM problem, Eq.~\ref{eq:sr_admm_l1l2l1_prob}, is then broken into the
following sub-problems for the two unknowns $\xx$ and $\zz$.
\begin{subequations}
  \begin{align}
    \xx^{(k+1)}  =& \scalebox{0.95}{$\argmin_{\xx} \Le_{\pel}(\xx, \zz^{(k)}, \ww^{(k)})$}
                    \label{eq:sr:l1l2l1_gen_admm_step_x}\\
    =& \scalebox{0.95}{$\argmin_{\xx} \norm{A\xx - \bb}_2^2 + \frac{\pel}{2}\norm{F\xx - \zz^{(k)} -\bb'+  \frac{\ww^{(k)}}{\pel}}_2^2$}
       \nonumber\\
    \zz^{(k+1)}  =& \scalebox{0.95}{$\argmin_{\xx} \Le_{\pel}(\xx^{(k+1)}, \zz, \ww^{(k)})$}
                    \label{eq:sr:l1l2l1_gen_admm_step_z}\\
    =& \scalebox{0.95}{$\argmin_{\zz} \norm{\zz}_1 + \frac{\pel}{2} \norm{ \zz - \left(  F\xx^{(k+1)} -\bb'+ \frac{\ww^{(k)}}{\pel}\right)}_2^2$}
       \nonumber\\
    \ww^{(k+1)} =& \scalebox{0.95}{$\ww^{(k)} + \pel\big( F\xx^{(k+1)} - \zz^{(k+1)} -\bb' \big)$}
                   \label{eq:sr:l1l2l1_gen_admm_step_w}
  \end{align}
  \label{eq:sr:l1l2l1_gen_admm_steps}
\end{subequations}
The sub-problem of $\zz$, in Eq.~\ref{eq:sr:l1l2l1_gen_admm_step_z}, is actually
a proximal operator of $\ell^1$ function,
\[
  \zz^{(k+1)} = \prox_{\pel^{-1}\norm{\cdot}_1}\left( F\xx^{(k+1)} - \bb' + \frac{\ww^{(k)}}{\pel} \right),
\]
which has the following closed form solution
\begin{equation}
  \begin{split}
    \zz^{(k+1)} =&\left[ \abs{F\xx^{(k+1)} - \bb' + \frac{\ww^{(k)}}{\pel}}  - \frac{1}{\pel}\right]_{+}\\
    &\quad\odot \sgn\left(  F\xx^{(k+1)} - \bb' + \frac{\ww^{(k)}}{\pel} \right)
  \end{split}
  \label{eq:sr_l1l2l1_prox}
\end{equation}

The sub-problem of $\xx$ (Eq.~\ref{eq:sr:l1l2l1_gen_admm_step_x}) has the
form of a least square approximation problem,
\begin{equation}
  \begin{split}
    \tilde{\xx}= \argmin\limits_{\xx} \norm{G\xx-\ccc}_2^2  \end{split},
  \label{eq:sr_l1l2l1_lsf}
\end{equation}
with \scalebox{0.88}{$G=\begin{bmatrix} A\\ \sqrt{\vartheta/2}F \end{bmatrix}$},
and  \scalebox{0.88}{$\ccc=\begin{bmatrix}
    \bb\\ \sqrt{\vartheta/2}\left( \bz^{(k)} + \bb' - \bw^{(k)}/\vartheta\right)
  \end{bmatrix}$}.
Equation~\ref{eq:sr_l1l2l1_lsf} can be effectively solved with a conjugate
gradient approach on normal equation~\cite{Boyd2004convex}.
\subsection{Treatment of Linear Operators}
\label{sec:sr_treat}
All computations are eventually broken down to matrix multiplication for which the largest computational efforts are on $A_k, S_d,\quad k\in[1,s_k], d\in[1,s_d]$, and their adjoint versions $A_k^\T, S_d^\T$.
These matrixes are very large and sparse.
For example, given a pair of low-resolution and high-resolution: $s_x\times s_y:= 128\times 128$ and $s_X\times s_Y=512\times 512$ (i.e., $4\times $ super-resolution).
Assuming $s_k= 16$ and $s_d = 8$, the size of $A$ is $2^{18}\times 2^{18}$ and the size of $S$ is  $2^{21}\times 2^{18}$.
Direct computation of these matrixes is infeasible.
Therefore, we decided to implement these matrices in the form of linear functions of 2D variables instead of sparse matrix and vectorized inputs.
\begin{figure}
  \centering
  \includegraphics[width=0.8\linewidth]{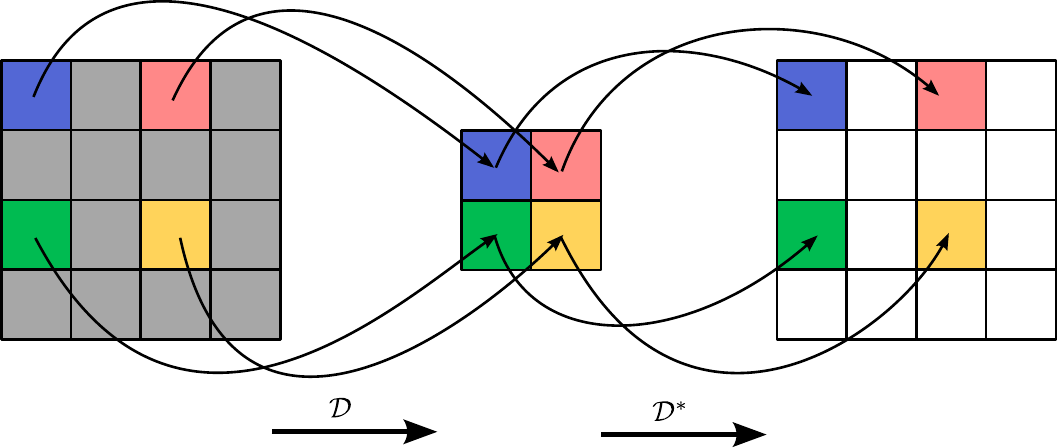}
  \caption[]{Implementation of downsampling operator}
  \label{fig:sr_down_op}
\end{figure}

For downsampling operator $\opD$, a simple resampling scheme is employed as depicted in Fig.~\ref{fig:sr_down_op}. For each block of $\Scale_x\times \Scale_y$ pixels, one pixel at the top-left location is picked and put into the low-resolution grid.
The adjoint operator $\opD^{*}$ is therefore simply putting back the corresponding pixel to this location.
The bluring operator $\opB$ is modelled by a simple Gaussian kernel with a standard deviation of $\sigma=\frac{1}{4}\sqrt{\Scale^2-1}$ and a size of $3\sigma$ as suggested in~\cite{Unger2010convex}.
The warping operator $\opW_k$ and its adjoint operator $\opW_k^{*}$ are implemented as forward-warping and backward-warping functions. These functions are associated with a set of disparity maps at each of the perspectives employed for super-resolution.
Assumes that a set of $s_k$ low-resolution sub-aperture images each with its perspective index is in $P=\{\Vs_1, \Vs_2,...,\Vs_{s_k}\}$ are inputs to estimate an super-resoltion image at $\Vs_0\in P$.
For each perspective $\Vs_k$, we need to find the disparity map $\Dz_k$.
The forward warping function $\opW_k$ will warp the SAI from perspective $\Vs_0$ to  $\Vs_k$ using $\Dz_k$, i.e., $\wLf(\Xy,\Vs_k) = \Lf(\Xy + \Vs_k\Dz_k,\Vs_k)$, while the backward warping function $\opW_k^{*}$ will warp the input SAI from perspective $\Vs_k$ to $\Vs_0$ using $\Dz_0$, i.e., $\wLf(\Xy,\Vs_0) = \Lf(\Xy + \Vs_0\Dz_0,\Vs_0)$.

The transformation matrix $S$ can be implemented in the form of weighted directional gradient ($\nabla^{U,V}$) computed for a direction set $ U=\{\bd_i|\bd_i\in \mathbb{N}^2, i=1,..,s_d\}$ and a weight set $V=\{V_i|V_i\in \mathbb{R}^{s_X\Stimes s_Y}, i=1,..,s_d\}$.
Let $I$ be the SAI at perspective $\Vs_0$, $I(\Xy) = \Lf(\Xy,\Vs_0)$, we computed $\nabla^{U,V}I$ as follow,
\begin{equation}
  \boldsymbol{G} = \nabla^{U,V}I = \left(\frac{\partial}{\partial \bd_1},
    \frac{\partial}{\partial \bd_2}, .., \frac{\partial}{\partial
      \bd_{s_d}}\right) I,
\end{equation}
with the weighted directional derivative $\partial/\partial \bd_i$ approximated by finite differences,
\begin{equation}
  \boldsymbol{G}_{d_i}(\Xy) =  \frac{\partial}{\partial \bd_i} I(\Xy) = V_i(\Xy) \left( I(\Xy) - I(\Xy+\bd_i)\right).
\end{equation}
The adjoint matrix $S^\T$ is then computed in the form of weighted directional
divergence,
\begin{equation}
  \texttt{div}^{U,V}\boldsymbol{G} = \nabla^{U,V}\cdot \boldsymbol{G} =
  \sum_{i=1}^{s_d} \frac{\partial \boldsymbol{G}_{\bd_i}}{\partial \bd_i}.
\end{equation}
\section{GPU-Accelerated Architecture}
\label{sec:sr_gpu_architecture}
This section presents the accelerated architecture for 4D LFSR.
Acceleration is achieved by parallel computation on graphics processing units.
Due to the multi-platform compatibility, we select OpenCL over CUDA for the
implementation of the proposed approach.
To solve the cost function optimization problem of Eq.~\ref{eq:sr_fin}, we
follow the iterative solving process discussed in Sec.~\ref{sec:sr_opt_ap}.
As will be discussed later in the experimental results (Sec.~\ref{sec:sr_exp_lfsr}), the ADMM solver provides better performance in optimizing the cost function as compared to the gradient descent approach.
\SetAlgoSkip{SkipBeforeAndAftera}
\begin{algorithm}
  \SetAlgoLined
  \DontPrintSemicolon
  \KwIn{$\vartheta,\xx_0, N$}
  \KwOut{$\xx$}
  $\xx^{(0)} := \xx_0$\;
  $\ww^{(0)} := 0 $\;
  \For{$n$ {\bf in} $1,2,..N$}{
    $\ba := A\xx^{(n-1)} - \bb$\;
    $\bu := F\xx^{(n-1)} - \bb' + \ww^{(n-1)}$\;
    $\zz^{(n)} := \prox_{\rho^{-1}\norm{\cdot}_1}(\bu)$
    \Comment*[r]{\small Solving Eq.~\ref{eq:sr:l1l2l1_gen_admm_step_z}}
    $\ww^{(n)} := \bu -\zz^{(n)}$
    \Comment*[r]{\small Computing Eq.~\ref{eq:sr:l1l2l1_gen_admm_step_w}}
    $\ff := 2\ww^{(n)}-\ww^{(n-1)}$\;
    $\bv := A^\T\ba + \frac{\vartheta}{2} F^\T\ff$\;
    $\xx^{(n)} := \texttt{xstep}(\bv,\xx^{(n-1)})$
    \Comment*[r]{\small Solving Eq.~\ref{eq:sr:l1l2l1_gen_admm_step_x}}
  }
  \KwRet $\xx^{(n)}$\;
  \caption{Minimization of the cost function in Eq.~\ref{eq:sr_fin} with ADMM
    iterative solver.}
  \label{alg:sr_admm_l2l1l1_overall}
\end{algorithm}

\begin{figure}
  \centering
  \includegraphics[width=\linewidth]{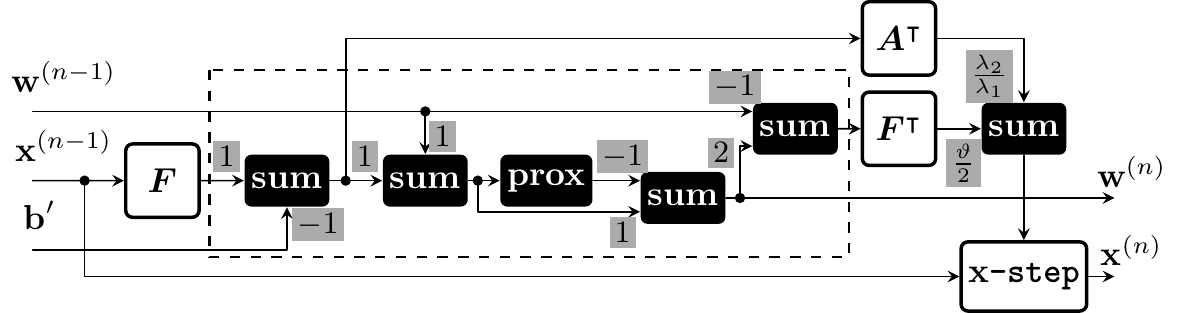}
  \captionsetup{belowskip=-15pt}
  \caption[]{Computation flow of one ADMM iteration.}
  \label{fig:sr_gpu_admm}
\end{figure}

For a better handling of the computation flow, we did the following
modifications to ADMM iteration in Eq.~\ref{eq:sr:l1l2l1_gen_admm_steps}.
First, the order of sub-problems is rearranged such that $\xx$-step comes after $\bz$-step and $\bw$-step.
This way allows us to make use of the computation of $F\xx$ for all sub-problems.
Secondly, the parameter $\vartheta$ is absorbed into $\bw$ (i.e., $\bw$ instead of $\bw/\vartheta$) to save unnecessary scalar multiplications. $\vartheta$ only takes part in the computation of proximal operator ($z$-step) and solving of the least square problem ($x$-step).
Fig.~\ref{fig:sr_gpu_admm} illustrates the modified computations of ADMM solver which is also listed in Algorithm~\ref{alg:sr_admm_l2l1l1_overall}.

The ADMM solver takes in three arguments, the parameter $\vartheta$, an initial
guess ($\xx_0$) and the number of iterations ($N$), as in
Algorithm~\ref{alg:sr_admm_l2l1l1_overall}.
Before the iteration, we initialized $\xx$ with $\xx_0$, a bi-cubic up-sampling of the low-resolution image, and $\bw$ with zeros, line $1,2$.
Each iteration starts with the computation of $A\xx$ and $F\xx$ which are associated to $\ell^2$ and $\ell^1$ terms of the objective function, Eq.~\ref{eq:sr_l1l2l1_obj}.
While $A\xx$ is subtracted by $\bb$, line $4$, $F\xx$ is subtracted by $\bb'$ and summed with $\bw$, line $5$.
Since $F$ is a stack of $A$ and $S$ and $\bb'$ includes $\bb$,
Eq.~\ref{eq:sr_admm_compact}, we avoid the re-computation of $A\xx-\bb$ by
extracting it from $F\xx-\bb'$ as depicted in Fig.~\ref{fig:sr_gpu_admm}.
The sum and subtract operations in line $5$ are realized by two-arguments sum kernels (i.e., $\texttt{sum}$ in Fig.~\ref{fig:sr_gpu_admm}).
The gray box attached to each input to the sum kernel denotes the scalar scaling of the input.
On line $6$, we conduct a $\zz$-step by computing the proximal operator of $\bu$.
This proximal operator is realized by an OpenCL kernel $\texttt{prox}$, as in Fig.~\ref{fig:sr_gpu_admm}, followed by a sum kernel which realizes $\ww$-step, line $7$ Algorithm~\ref{alg:sr_admm_l2l1l1_overall}.

After the computation of $\zz$ and $\ww$, the next step is preparing the residual input for the conjugate gradient descent solver in $\xx$-step,
$\bv = G^\T(G\xx -\ccc)$. From Eq.~\ref{eq:sr_l1l2l1_lsf}, we have
\begin{gather+}[0.9]
  \bv =&\quad \begin{bmatrix}A\\ \sqrt{\vartheta/2}F\end{bmatrix}^T
  \begin{bmatrix}A\xx-\bb\\\sqrt{\vartheta/2}\left(F\xx- \zz^{(n)}-\bb'+\ww^{(n)} \right)\end{bmatrix} \\
  =&\quad A^\T(A\xx-\bb) + \vartheta/2 F^\T (F\xx-\zz^{(n)}-\bb'+\ww^{(n)})\\
  =&\quad A^\T\ba + \frac{\vartheta}{2} F^T\ff
\end{gather+}

With the computation of $\ff$, Algorithm~\ref{alg:sr_admm_l2l1l1_overall} line
$8$, as
$\ff = 2 \ww^{(n)} - \ww^{(n-1)} = \uu -\zz - \ww^{(n-1)} + \ww^{(n)} =
F\xx^{(n-1)}-\zz - \bb'+\ww^{(n)}$.
The computations of $\ff$ and $\bv$ are realized by two sum kernels directly before and after $F^{T}$ as in Fig.~\ref{fig:sr_gpu_admm}. Notice that we made a scaling of $A^T\ba$ by $\frac{\lambda_2}{\lambda_1}$ since $\ba$ is extracted from $F\xx-\bb'$ which has a different scalar scaling of matrix $A$ and column vector $\bb$. Another note from the implementation of
Fig.~\ref{fig:sr_gpu_admm} is that the group of OpenCL kernels marked by dashed rectangle would be combined into a single kernel, since these kernels share element-wise operators.

\begin{figure}
  \centering
  \includegraphics[width=0.9\linewidth]{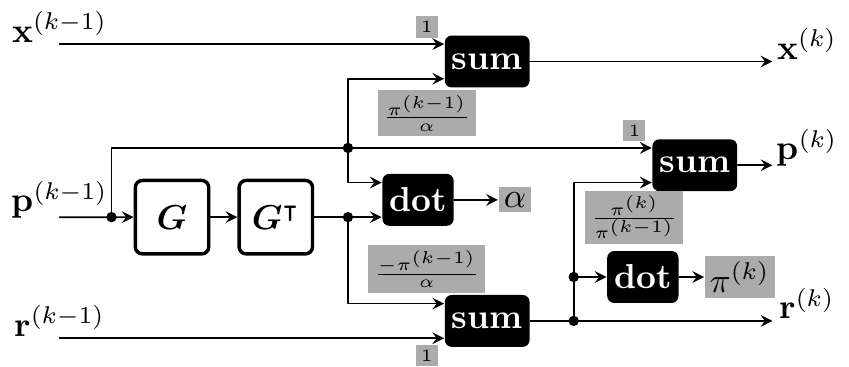}
  \captionsetup{belowskip=-15pt}
  \caption[]{Computation flow of $\texttt{xstep}$.}
  \label{fig:sr_gpu_xstep}
\end{figure}
As discussed in the previous section, conjugate gradient descent on normal equation is employed to solve $\ell^2$ optimization problem of $\xx$-step.
Fig.~\ref{fig:sr_gpu_xstep} depicts the computation flow of $\xx$-step, while its pseudo code is listed in Algorithm~\ref{alg:sr_admm_cg}.
There are two inputs, i.e. $\bv$, $\xx^{(n-1)}$, and two scalar parameters, i.e., $\tau$, $K$.
The computed HR image from the previous ADMM iteration $\xx^{(n-1)}$ is used as
the initial guess for the conjugate gradient descent solver, while the residual
$\bv$ is used to initialize $\br^{(0)}$, $\bp^{0}$ and compute the initial error
$\pi^{(0)}$.
The two parameters $\tau$ and $K$ specify the error threshold and the maximum number of conjugate gradient iterations, respectively.
The stop condition is that either the residual $\br$ is sufficiently small or the maximum number of iterations is reached, Algorithm~\ref{alg:sr_admm_cg} line $4$,$5$.
All computations in Algorithm~\ref{alg:sr_admm_cg} can be effectively broken down into GPU kernel implementation.
Beside the forward and backward transform  ($G,G^\T$), there are two kernels $\texttt{sum}$ and $\texttt{dot}$, as in Fig.~\ref{fig:sr_gpu_xstep}, which represents element-wise sum and dot product respectively.

\begin{figure}
  \centering
  \includegraphics[width=0.8\linewidth]{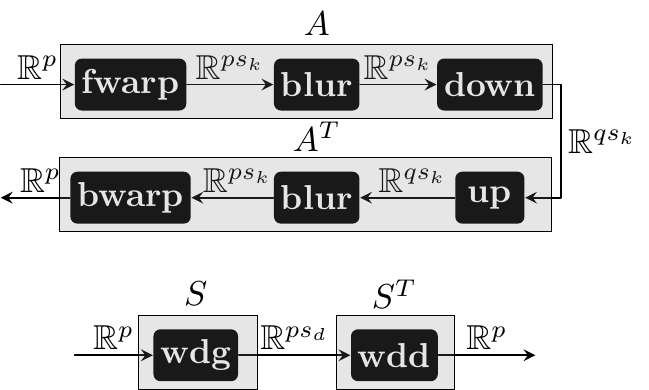}
  \captionsetup{belowskip=-15pt}
  \caption[]{Kernel realization of $A,A^\T,S,S^\T$}
  \label{fig:sr_gpu_admm_as}
\end{figure}

From the Eq.~\ref{eq:sr_admm_compact} and Eq.~\ref{eq:sr_l1l2l1_lsf}, we can
derive the computation of $G^TG$ in the form of $A$ and $S$ as
\begin{equation}
  \small
  \begin{split}
    G^TG = A^TA + \frac{\vartheta}{2}F^\T F =  \left(1+\frac{\vartheta}{2}\frac{\alpha_1^2}{\alpha_2}\right)A^TA + \frac{\vartheta}{2}S^TS,
  \end{split}
\end{equation}
with the kernel realization of $A$,$S$ and its adjoint version $A^\T$, $S^\T$ shown in Fig.~\ref{fig:sr_gpu_admm_as}.
The figure illustrates the change in the size of the column vector after each kernel execution.
Regarding Fig.~\ref{fig:sr_gpu_admm_as}, \texttt{fwarp}, \texttt{bwarp}, \texttt{blur}, \texttt{up}, and \texttt{down} denote the forward warp, backward warp, blur,  up-sampling and down-sampling kernel respectively. \texttt{wdg} kernel realizes the weighted directional gradient (i.e., $\nabla^{U,V}$), while the weighted directional divergence (i.e., $\texttt{div}^{U,V}$) is implemented by \texttt{wdd} kernel.

\SetAlgoSkip{SkipBeforeAndAfterb}
\begin{algorithm}
  \SetAlgoLined
  \DontPrintSemicolon
  \KwIn{$\bv, \xx^{(n-1)},\tau, K$}
  \KwOut{$\xx$}
  $\xx^{(0)} := \xx^{(n-1)}$\;
  $\bp^{(0)} := \br^{(0)}:= \bv $\;
  $\pi^{(0)}:= <\br^{(0)},\br^{(0)}>$
  \Comment*[r]{\small dot product}
  \For{$k$ {\bf in} $1,2,..K$}{
    \If{$\pi^{(k-1)} < \tau$}{
      $\bv := G^{\T}G\bp^{(k-1)}$\;
      $\alpha := <\bv,\bp^{(k-1)}>$\;
      $\br^{(k)} :=\br^{(k-1)} - \frac{\pi^{(k-1)}}{\alpha}\bv^{(k-1)}$\;
      $\pi^{(k)} := <\br^{(k)},\br^{(k)}>$\;
      $\bp^{(k)} := \bp^{(k-1)} + \frac{\pi^{(k)}}{\pi^{(k-1)}}\br^{(k-1)}$\;
      $\xx^{(k)}:= \xx^{(k-1)} + \frac{\pi^{(k-1)}}{\alpha}\bp^{(k-1)}$\;
    }
  }
  \KwRet $\xx^{(k)}$\;
  \caption{Solving $\xx$-step}
  \label{alg:sr_admm_cg}
\end{algorithm}

\section{Limitation and Discussion}
Although the strategy to realize the degradation process with linear functions has the advantage of saving computational resources and simplifying the GPU implementation, it presents a drawback in dealing with a more challenging blurring process, i.e., space-variant PSFs. In this work, we assume that the PSF is space-invariant and can be approximated by a single Gaussian blur kernel. However, depending on the optical setup, the blurring process may involve a set of space-variant PSFs. This means that each blur kernel may only be applied to a group of pixels, and different regions of an image would require different blur kernels. In such a case, sparse matrix realization of blurring operator would be a reasonable option to avoid the complication of maintaining and applying region-specific blur kernel.

Beside Gaussian and Impulse noise, there is another challenging noise originating from the discrete nature of the electric charge, namely photon noise or shot noise \cite{Hasinoff2014photon}.
Different from additive Gaussian noise, which is pixel independent, the photon noise is pixel dependent and follows the Poisson distribution. Taking the notation from Sec.~\ref{sec:sr_degrad}, the degradation model considering Poisson noise and additive Gaussian noise reads
\begin{equation}
  \yy_k = \zz_k + \bepsilon,
\end{equation}
where $\zz_k \sim \opP(A_k\xx)$ and
$\bepsilon \sim \opN(0,\sigma^2)$ represent the Poisson distribution and a
zero-mean Gaussian distribution, respectively.
Following the work in \cite{Sun2020multi}, we can rewrite our data fidelity term
as
\begin{gather+}[0.9]
  \label{eq:sr_opt_poi}
  E(\xx) = \sum\limits_{k=1}^{s_k}\norm{A_k\xx-\yy_k}_{W_k}^2 + <log\big(A_k\xx
  + \sigma^2\big),1>,
\end{gather+}
where $log(\cdot)$ is computed element-wise and diagonal weight matrix $W_i$ is computed as
\begin{equation}
  W_k = diag\left( \frac{1}{[A_k\xx]_i + [\sigma]_i^2} \right),
\end{equation}
with $[\xx]_i$ denotes the $i^{th}$ element of column vector $\xx$.
As discussed in \cite{Sun2020multi}, although $\ell 1$/$\ell 2$ data terms can also be applied to input data with Poisson noise, their reconstruction quality is about 1dB worse as compared to applying Eq.~\ref{eq:sr_opt_poi}. Due to the $log$ function, the above data term will lead to a non-convex optimization problem in which a global minimum is not guaranteed. For solving this new problem, a new decomposing strategy with ADMM needs to be developed. This task, together with the acceleration of the new solving process, is listed in our plan for future work.
\section{Experimental Results}
\label{sec:sr_experiment}
This section discusses the results of our experiments, in which the robustness of the proposed SR model is validated through numerous testing scenarios.
Comparisons to the state-of-the-art approaches under severe mixed noise conditions and previous GPU acceleration approaches are presented. In addition,  the performance of the accelerated computational framework is also analyzed and discussed.
\subsection{Evaluation of LFSR Computational Framework}
\label{sec:sr_exp_lfsr}
Light-field scenes from 4D synthetic dataset~\cite{Honauer2016dataset} are employed to evaluate the robustness of the SR model and analyze the converge of iterative solvers.
This dataset is selected since it includes plenty of scenery and provides accurate disparity maps.
We follow the degradation model discussed in Sec.~\ref{sec:sr_degrad} to prepare the input data with two test scaling factors, i.e., $\Stimes 2$, $\Stimes 4$.
The observation noises are parameterized by $\sigma$ and $\nu$, which respectively denotes the standard deviation of Gaussian noise (i.e., $\mathcal{N}(\mu,\sigma)$) and the percentage of impulse noise (i.e., salt and pepper).
In order to match the practical use cases in which the high-resolution disparity maps are not available, the provided disparity maps are down-scaled by the same factor as of the test case (i.e., $\Stimes 2$, $\Stimes 4$) and then are interpolated back
to the original size and used in the warping functions.
For handling color input data, we follows the strategy proposed
in~\cite{Tran2018gpu} to solve the cost function for $Y$ color channel while
applying bi-cubic interpolation for $Cb$ and $Cr$ channel.
\begin{figure}[t]
  \centering
  \begin{minipage}[b]{\linewidth}
    \centering
    \includegraphics[width=0.9\textwidth]{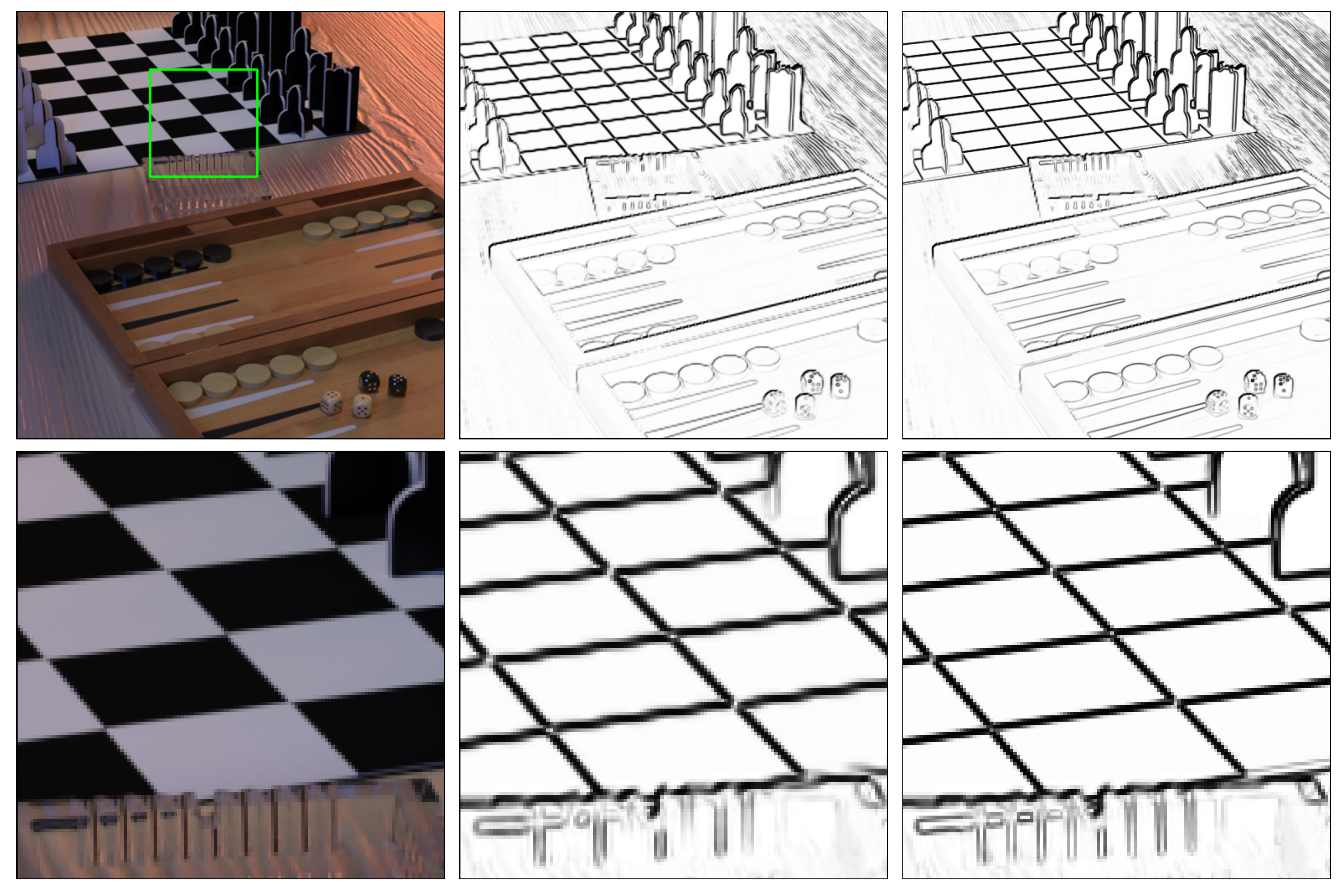}

    \vspace{-1.25em}
    \begin{flushleft}
      \hspace{5.1em}(a)
      \hspace{6.5em}(b)
      \hspace{6.5em}(c)
    \end{flushleft}
  \end{minipage}
  \caption{Regularization weights calculation for LF scene `\textit{boardgames}';
    top row: full size image and weights; bottom row: zoom-in of region marked
    by green rectangle; (a) Ground-truth image; (b) weights at the $1^{st}$ iteration;
    (c) weights after the $10^{th}$ iteration.}
  \label{fig:sr_weighting}
\end{figure}

\begin{figure}[t]
  \centering
  \begin{minipage}[b]{\linewidth}
    \centering
    \includegraphics[width=\textwidth]{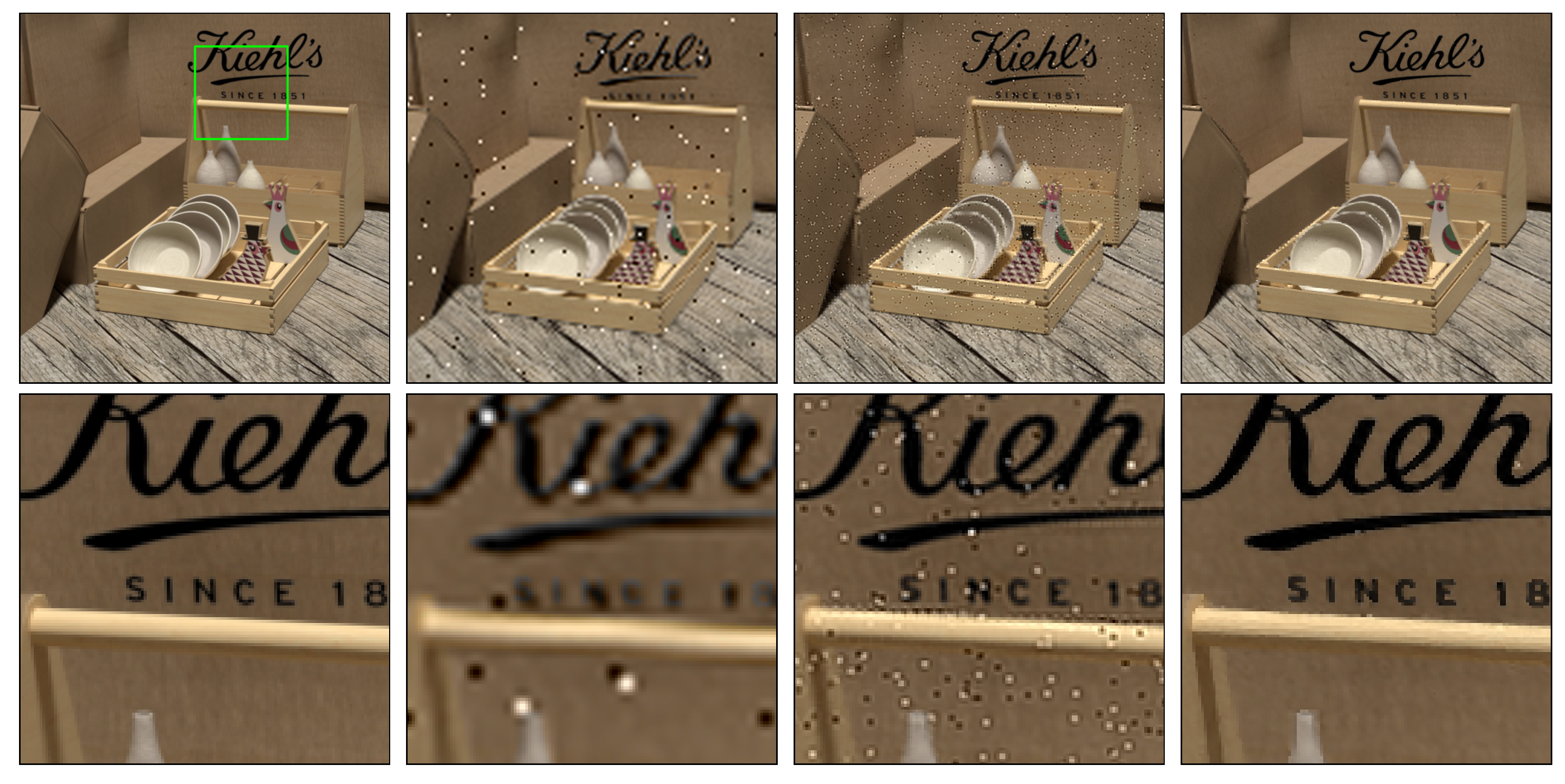}

    \vspace{-1.25em}
    \begin{flushleft}
      \hspace{3.5em}(a)
      \hspace{5.20em}(b)
      \hspace{5.20em}(c)
      \hspace{5.20em}(d)
    \end{flushleft}
  \end{minipage}
  \captionsetup{belowskip=-15pt}
  \caption{ $\times 4$ super-resolution of LF scene
    `\textit{dishes}' ($\sigma=1$, $\nu=1\%$);
    top row: full size image; bottom row: zoom-in of region marked
    by green rectangle; (a) Ground-truth image; (b) bi-cubic up-sampling (21.72 dB);
    (c) 1st iteration (24.96 dB)
    (d) 5th iteration (29.53 dB)}
  \label{fig:sr_sr_x4}
\end{figure}

The regularization weights computed for the scene `\textit{boardgames}' are
shown Fig.~\ref{fig:sr_weighting}.
It is expected that a strong weighting is applied to the region where
high-frequency information occupied (i.e., texture edges, occlusions). As
discussed in Sec.~\ref{sec:sr_opt_reg_term}, the regularization weight is a
combination of  spatial weight ($w_d$), edge weight ($\bw_e$), and occlusion
weight ($\bw_o$).
To strengthen the regularizing effect, the weights are recomputed for each ADMM
iteration using the current computed super-resolution image $\xx$.
When the optimization starts, $\xx$ is initialized to a bi-cubic up-sampling of
the low-resolution image.
This explains the blur edges of regularization weights at iteration 1, as shown
in Fig.~\ref{fig:sr_weighting} (b).
However, it could be observed after each ADMM iteration that the qualities of $\xx$ and regularization weight are gradually improved.
As shown in Fig.~\ref{fig:sr_weighting} (c), the regularization weight after 10
ADMM iterations capture well the high-resolution structure of the reconstructed
scene.

Fig.~\ref{fig:sr_sr_x4} visualizes the SR result for the $\Stimes 4$ test case of LF scene `\textit{dishes}'.
We employed 17 LR sub-aperture images as inputs to calculate the cost function in Eq.~\ref{eq:sr_fin} which is then solved by ADMM iterative solver.
The SAIs are picked up from $5\Stimes 5$ angular views in a star-like structure.
As compared to the bi-cubic up-sampling image used as an initial solution (Fig.~\ref{fig:sr_sr_x4} (b)), the reconstructed HR image after the first ADMM iteration (Fig.~\ref{fig:sr_sr_x4} (c)) demonstrates an obvious improvement in visibility.
Although the noise effect from the combination of multiple SAIs is still visible, it is possible to observe the texture content (i.e., small characters in the middle of the zoom-in region).
After 5 ADMM iterations, the noise effect is removed, resulting in a significant enhancement in visual quality with 4.6dB and 7.8dB improvement as compared to the 1st iteration's solution and the initial solution, respectively.
\begin{figure}[t]
  \centering
  \begin{minipage}[b]{\linewidth}
    \centering
    \includegraphics[width=0.9\textwidth]{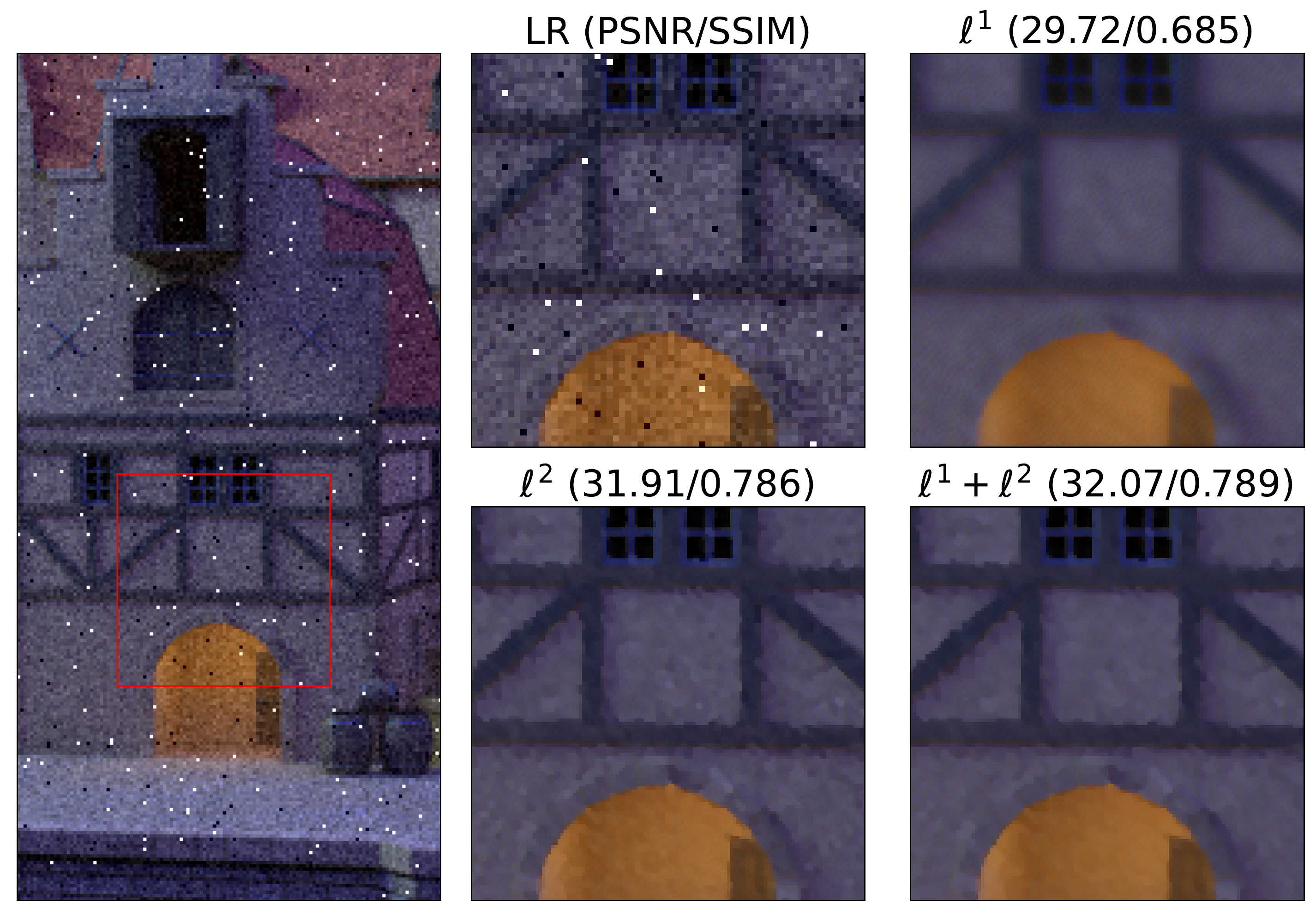}
  \end{minipage}
  \captionsetup{belowskip=-20pt}
  \caption{ $\times 2$ super-resolution of LF scene
    `\textit{medieval2}' ($\sigma=10, \nu=1\%$);
    \textit{left} : a cropped noisy LR input;
    \textit{right}: four zoom-in of the marked region from an LR input and three
    different configurations of data fidelity term.
  }
  \label{fig:sr_noise_x2}
\end{figure}

\begin{figure*}[t]
  \centering
  \begin{minipage}[b]{.9\linewidth}
    \centering
    \includegraphics[width=\textwidth]{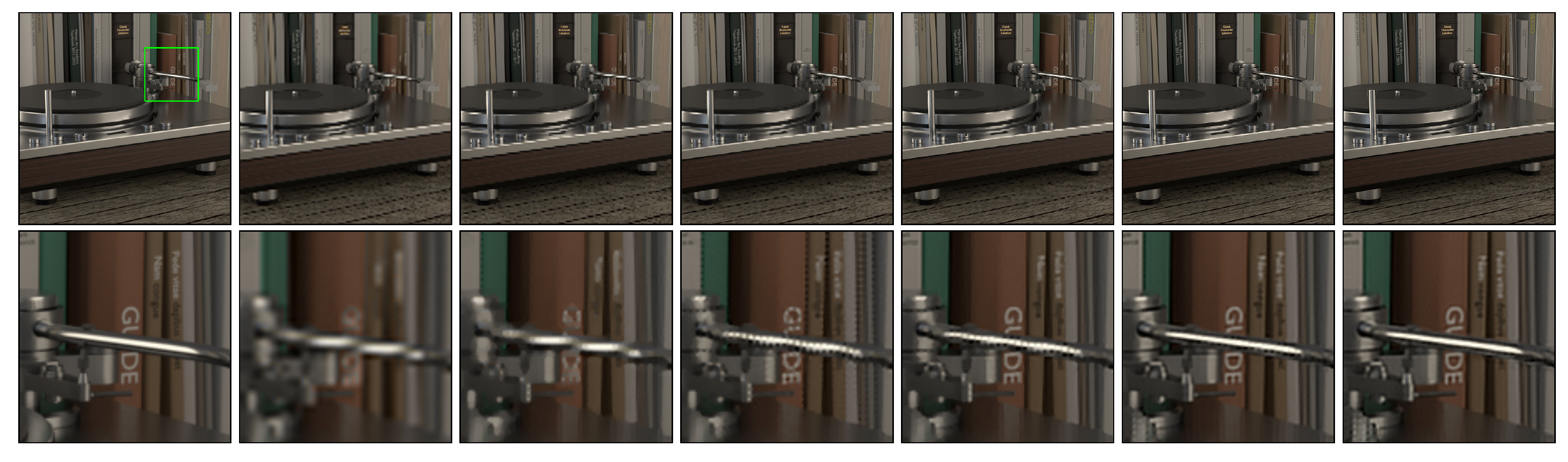}

    \vspace{-1.5em}
    \begin{flushleft}
      \hspace{3.5em}(a)
      \hspace{5.75em}(b)
      \hspace{5.75em}(c)
      \hspace{5.75em}(d)
      \hspace{5.75em}(e)
      \hspace{5.75em}(f)
      \hspace{5.75em}(g)
    \end{flushleft}
  \end{minipage}
  \captionsetup{belowskip=-15pt}
  \caption{Super-resolution $\times 4$ results of LF scene `\textit{vinyl}' under different
    number of inputs. \textit{top} full size HR image; \textit{bottom} zoom-in
    of marked region; (a) Ground-truth image; (b) bi-cubic upsampling (24.04dB);
    (c) 3 SAIs (27.55dB); (d) 5 SAIs (29.33dB); (e) 9 SAIs (30.54dB); (f)
    25 SAIs (31.65dB); (g) 49 SAIs (32.09dB).}
  \label{fig:sr_diff_input}
\end{figure*}

To evaluate the contribution of $\ell^1$ and $\ell^2$ data terms in reconstructing HR perspective image under mix-noise condition, we prepare a test case in which LR Light-field is severely damaged by noise effects, see Fig.~\ref{fig:sr_noise_x2}. While keeping the regularization part unchanged, we tuned data fidelity parameters ($\lambda_1$, $\lambda_2$) to find a solution with the highest PSNR score for each model (i.e., $\ell^1$, $\ell^2$, $\ell^1+\ell^2$).
We observed that using only $\ell^2$ data fidelity tends to oversmooth the solution due to the effect of the $\ell^2$ norm.
Although $\ell^1$ data fidelity well preserves the sharp edge structure, it also carries the effect of the noisy pixels into the solution.
The proposed mix-noise data term combines the impacts of both $\ell^2$ norm and $\ell^1$ norm and provides a better reconstruction quality.

The number of input LR images play an important role in the quality of reconstructed HR image.
Although demanding higher computation resources, we observed that more input SAIs tend to provide higher reconstruction qualities.
Fig.~\ref{fig:sr_diff_input} reports the $\times 4$ super-resolution results of LF scene `\textit{vinyl}' where different numbers of LR sub-aperture images are used.
As can be seen from the figure, giving more input images to the computational
problem (Eq.~\ref{eq:sr_fin}) results in the better visual quality of HR
solutions,
which is also evident from the reported PSNR scores.
Specifically, an improvement of 3.5dB as compared to bi-cubic up-sampling can be achieved with three input images.
When increasing the number of LR images to 5, 9, 25, and 49, we observed the incremental gains of 1.8dB, 1.2 dB, 1.1 dB and 0.44 dB, respectively.

\begin{figure}[t]
  \centering
  \begin{minipage}[b]{.9\linewidth}
    \centering
    \includegraphics[width=\textwidth]{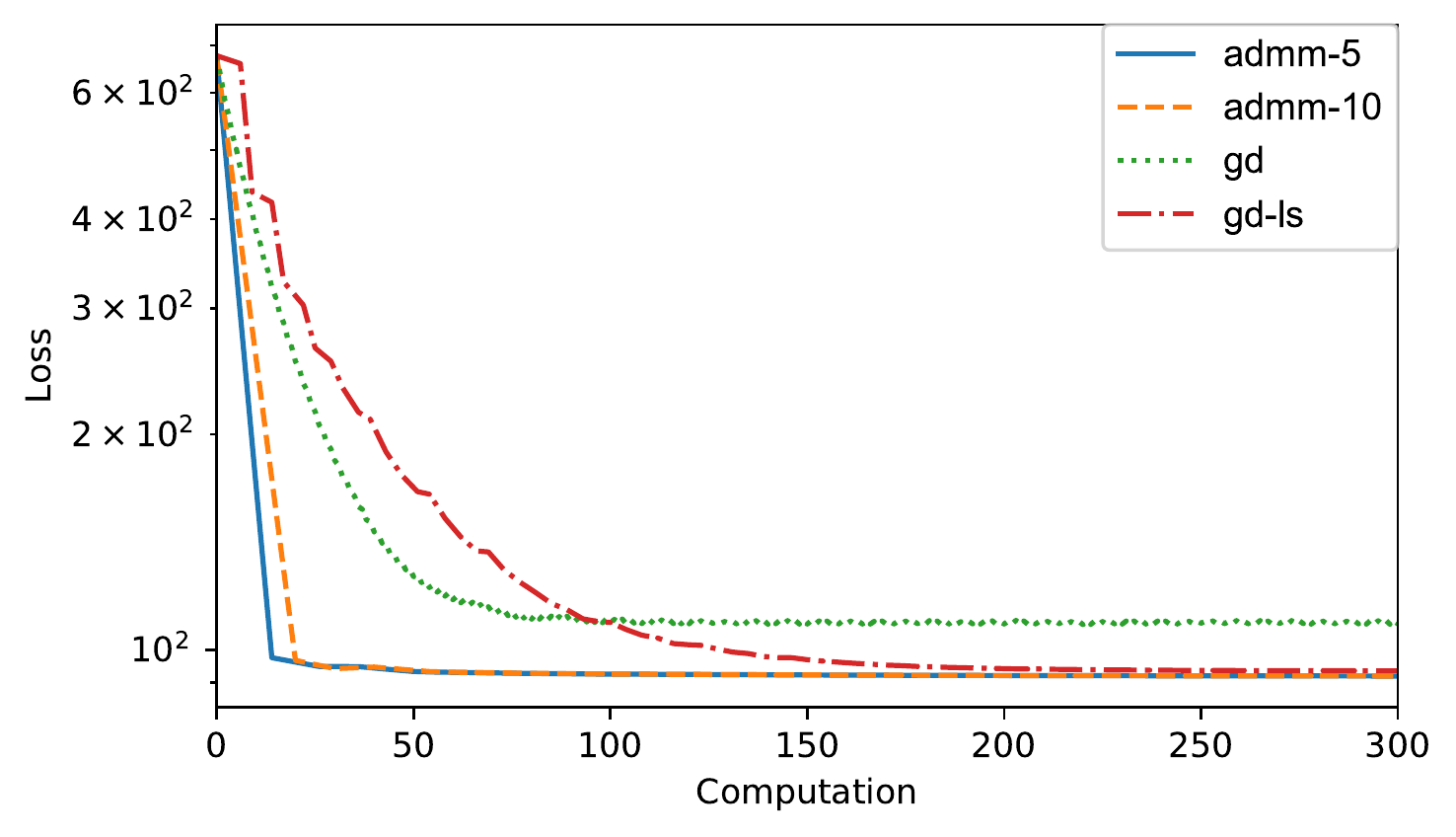}
  \end{minipage}
  \vspace{-10pt}
  \caption{Optimization results of different solvers.}
  \label{fig:sr_comp_py_solvers}
\end{figure}

In order to compare the convergence of the iterative solvers, we employ the matrix transform functions (i.e.,$A$,$S$) and their adjoint versions (i.e., $A^\T$, $S^\T$) as computation units (CU).
As derived in Sec.~\ref{sec:sr_opt_ap}, these transforms are the most dominant computation tasks and exist in every iterative step.
Each CU is either a combination of $A$ and $S$ as for computing the cost function $\tJ$ or $A^\T$ and $S^\T$ as for computing the gradient $\nabla\tJ$.
In this experiment, we built a cost function for $\times 2$ SR problem of LF scene `\textit{vinyl}' and applied four different configurations of the iterative solvers to optimize it. The first two are gradient descent solver (GD) without and with line search denoted as \textit{gd} and \textit{gd-ls } respectively.
The last two are ADMM solvers in which we configure the maximum number of conjugate gradient steps to 5 (\textit{admm-5}) and 10 (\textit{admm-10}).
Fig.~\ref{fig:sr_comp_py_solvers} presents the plot of the loss function against the accumulated CU.
Providing a good step size, GD without line search can make a rapid reduction in the cost function for the first few iterations.
However, due to fixed step size, the GD cannot optimize the loss function further after 80 CUs.
In contrast, \textit{gd-ls} seems slow at the beginning due to the search for an appropriate step size but is able to surpass
\textit{gd} at around 100 CUs and approach the global minimum after around 300 CUs.
Avoiding the costly line-search tasks, both configurations of the ADMM solver demonstrate a superior convergence rate as compared to GD.
We also observed that setting the maximum number of conjugate descent steps to 5 does shorten the computation effort for the first few iterations.
However, at later iterations when the early stop condition is satisfied, i.e., Algorithm~\ref{alg:sr_admm_cg} line \textit{5}, both settings result in a similar performance.

\begin{figure}[t]
  \centering
  \begin{minipage}[b]{\linewidth}
    \centering
    \includegraphics[width=\textwidth]{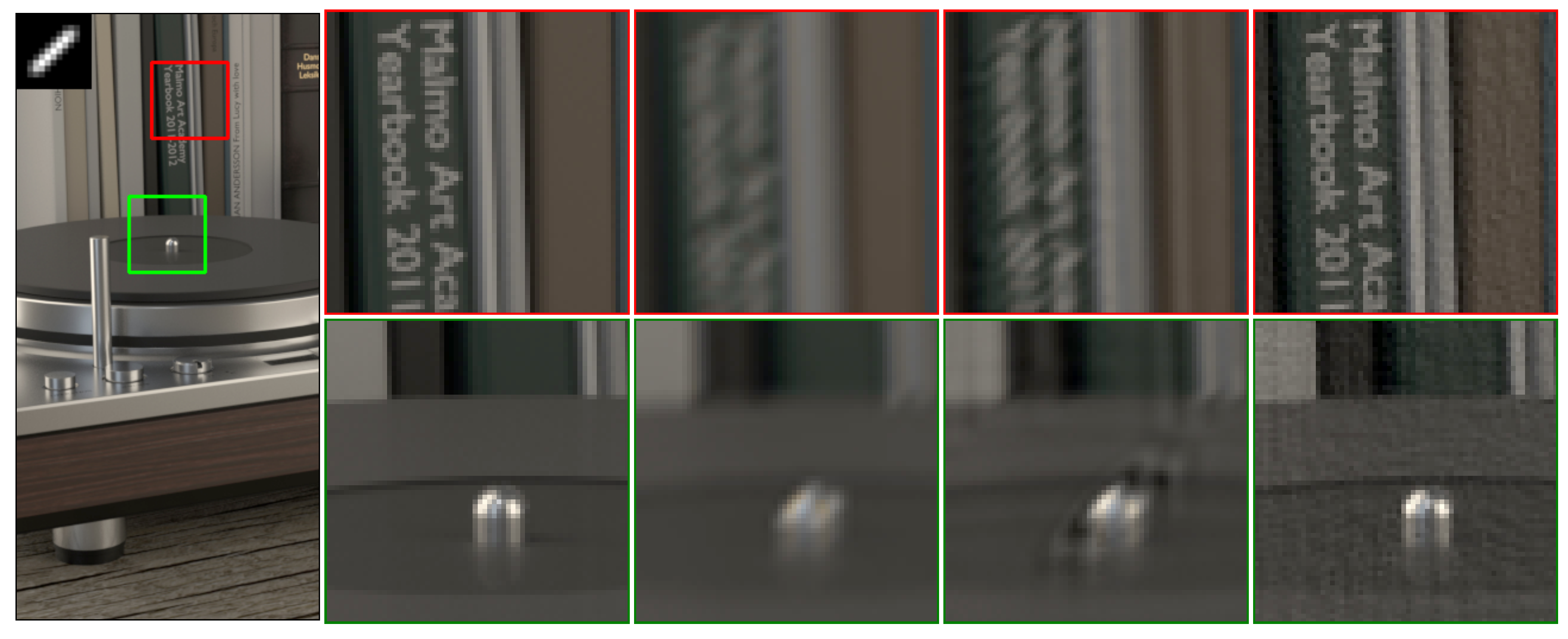}
  \end{minipage}
  \vspace{-18pt}
  \begin{flushleft}
    \hspace{2.25em} (a)
    \hspace{3.5em} (b)
    \hspace{3.5em} (c)
    \hspace{3.5em} (d)
    \hspace{3.5em} (e)
  \end{flushleft}
  \caption{$\Stimes 2$ super-resolution result of LF scene `vinyl' degraded by
    motion blur. (a) a cropped of ground truth with two marked region and motion
    blur kernel shown at top left corner; (b)
    zoom-in of ground truth image; (c)
    bi-cubic initial image (26.86dB); (d) after $1^{st}$ ADMM iteration
    (30.27dB);
    (e) after $10^{th}$ ADMM iteration (35.43dB).}
  \label{fig:sr_comp_motion_blur}
\end{figure}

The proposed computation framework can also be applied to a more challenging image condition such as motion blur. In such a case, the motion blur can be modelled by a convolutional kernel as a realization of the linear operator $\opB$ (see Fig.~\ref{fig:sr_degradation}). Fig.~\ref{fig:sr_comp_motion_blur} shows our $\Stimes 2$ SR result for low-resolution LF input degraded by a 45 degree motion blur. The blur kernel is shown on the top left corner of Fig.~\ref{fig:sr_comp_motion_blur}(a) and two zoom-in regions of the bi-cubic upsampling of degraded low-resolution SAI are shown in Fig.~\ref{fig:sr_comp_motion_blur}(c). Taking 25 SAIs as inputs to our reconstruction algorithm, we can achieve more than a 3 dB improvement in PSNR score after one ADMM iteration. The high-resolution LF image is well reconstructed after $10^{th}$ ADMM iterations with clear texture information and motion trace.
\begin{figure*}[t]
  \centering
  \begin{minipage}[b]{.9\linewidth}
    \centering
    \includegraphics[width=\textwidth]{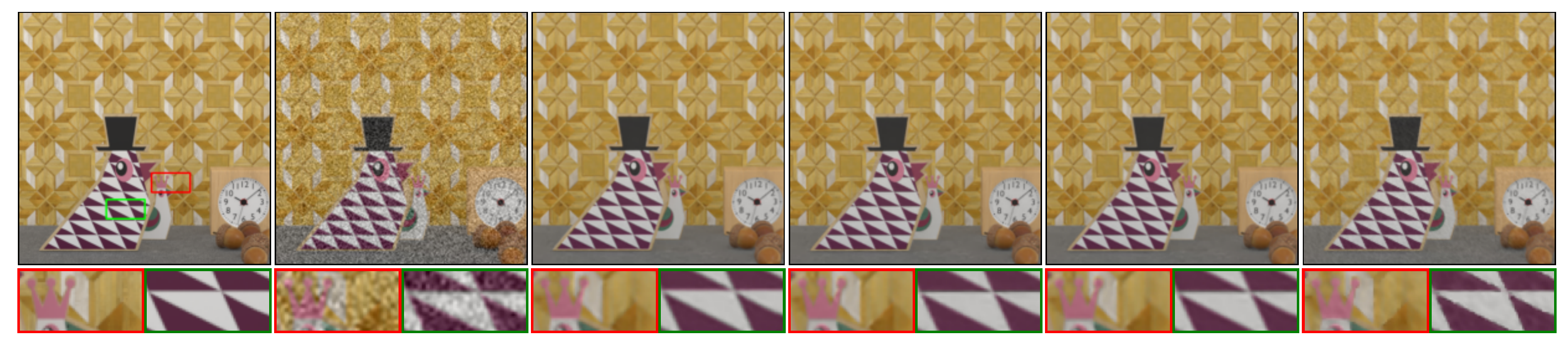}

    \vspace{-0.5em}
    \includegraphics[width=\textwidth]{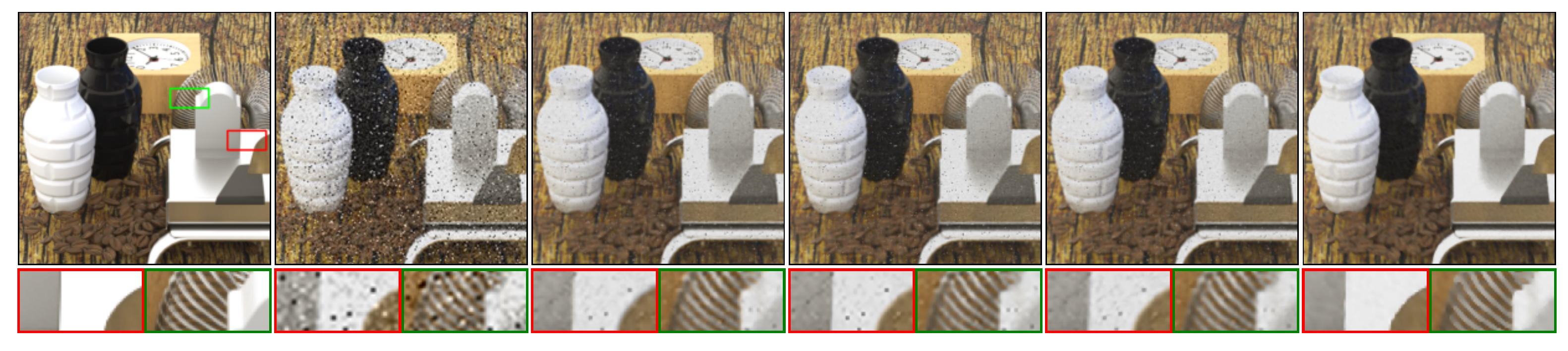}

    \vspace{-0.5em}
    \includegraphics[width=\textwidth]{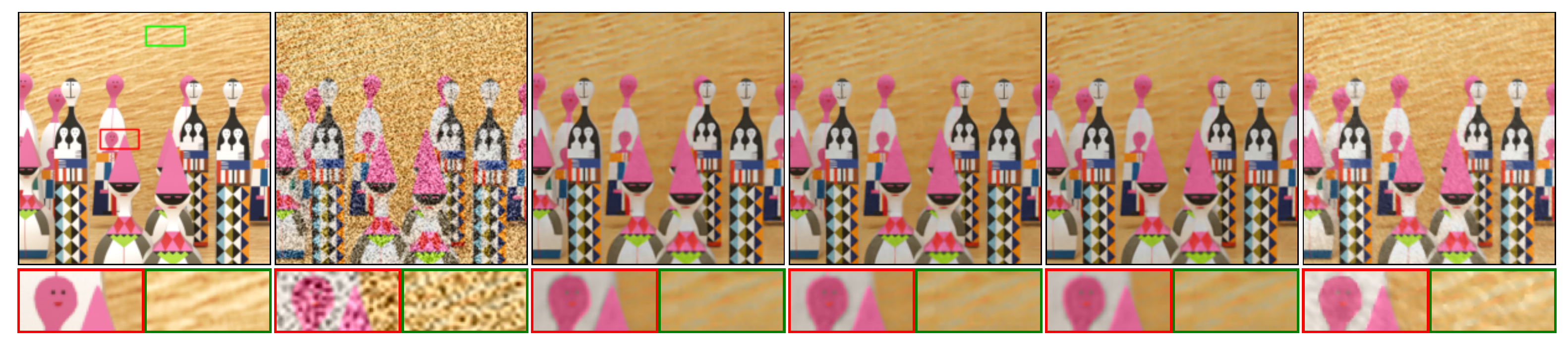}

    \vspace{-0.5em}
    \includegraphics[width=\textwidth]{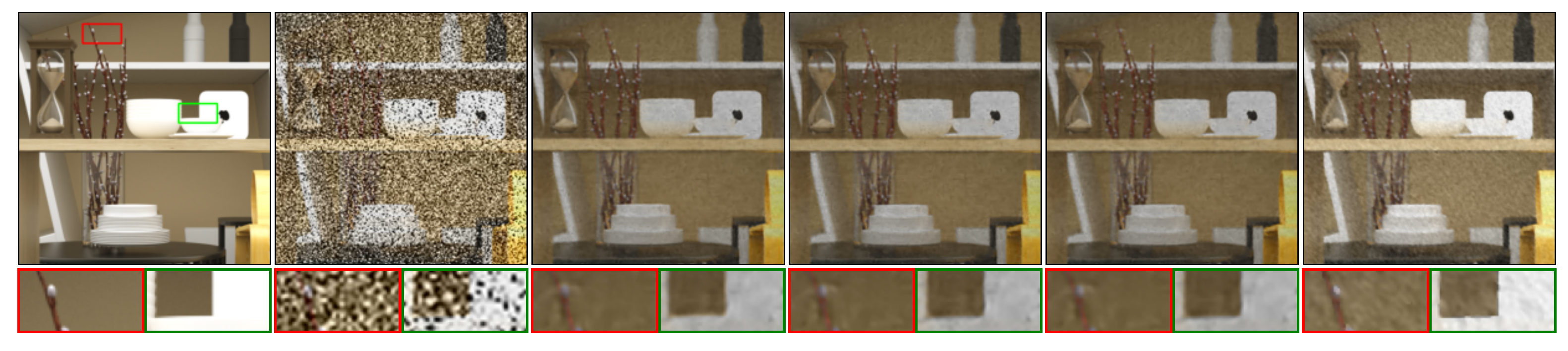}

    \vspace{-1.5em}
    \begin{flushleft}
      \hspace{1.5em} Ground Truth
      \hspace{2.25em} Input SAI
      \hspace{3.5em} De-resLF
      \hspace{3.5em} De-3DVSR
      \hspace{3.75em} DRLF
      \hspace{5.10em} Ours
    \end{flushleft}
  \end{minipage}
  \captionsetup{belowskip=-15pt}
  \caption{Visual comparisons of LFSR approaches under various
    mixed noise settings. From top to bottom (`scene' - $\sigma$/$\nu$):
    `Rooster-clock' - 20/0;
    `Coffee-beans-vases' - 20/5;
    `Smiling-crowd' - 50/0;
    `Dishes' - 50/20.
  }
  \label{fig:sr_x2_sta}
\end{figure*}
\subsection{Comparison to LFSR Approaches}
In this section, we evaluate the performance of the proposed method under severe
mixed noise conditions and compare it to state-of-the-art approaches (i.e., resLF~\cite{Zhang2019residual}, DRLF~\cite{Guo2021deep}, and 3DVSR~\cite{Tran20223dvsr}). These approaches currently provide state-of-the-art performance in reconstructing high-resolution LF images. To the best of our knowledge, only DRLF~\cite{Guo2021deep} supports LFSR with noisy input. For the evaluation, we randomly select five scenes from the Inria LF dataset~\cite{Shi2019framework}.
For each scene we generate low resolution LF ($\Stimes 2$) and insert noises with four
configurations, ($\sigma\texttt{=}20,\nu\texttt{=}0\%$), ($\sigma\texttt{=}20,\nu\texttt{=}5\%$),
($\sigma\texttt{=}50$, $\nu\texttt{=}0\%$), ($\sigma\texttt{=}50,\nu\texttt{=}20\%$).
These Gaussian noise settings are selected due to the pre-trained weights published by DRLF. DRLF needs different trainings for dealing with different noise conditions, and there are only three pre-trained weights published for three Gaussian noise configurations
$\sigma\texttt{=}10$, $\sigma\texttt{=}20$, and $\sigma\texttt{=}50$.
In addition, DRLF does not directly process noisy LR inputs. It provides separate networks for de-nosing and super-resolution. Therefore, we applied first their de-noising network to noisy LR inputs and then applied their SR network to the de-noised LR outputs. In this way, we are able to evaluate the performance of the other two state-of-the-art LFSR approaches (i.e., resLF~\cite{Zhang2019residual}, 3DVSR~\cite{Tran20223dvsr}) using the de-noised LR output from DRLF.

\begin{table*}[!h]
  \renewcommand{\arraystretch}{1.1}
  \centering
  \setlength{\tabcolsep}{2pt}
  \captionsetup{justification=centering}
  \begin{threeparttable}
    \vspace{8pt}
    \caption{Quantitative comparison of LFSR approaches under various mixed noise settings.}
    \begin{tabular}{ | c | c | c | c | c | c | c | c | c | }
      \hline
      Noise & Scenes &  BIC & resLF & De+resLF & 3DVSR & De+3DVSR & DRLF & Ours \\
      ($\sigma/\nu$) & & (psnr/ssim)& (psnr/ssim)& (psnr/ssim)& (psnr/ssim)& (psnr/ssim)& (psnr/ssim)& (psnr/ssim)\\
      \hline
      \hline
      \multirow{6}{*}{20/0}& Dishes& 22.63/0.378& 21.20/0.316& 28.80/0.886& 19.32/0.252& 28.74/0.891& 28.71/0.888& 30.47/0.846 \\
      \cline{2-9}
            & Rooster-clock& 22.81/0.375& 21.24/0.304& 30.73/0.864& 19.36/0.241& 31.15/0.879& 30.84/0.881& 31.48/0.798 \\
      \cline{2-9}
            & Coffee-beans-vases& 21.55/0.507& 20.41/0.453& 25.24/0.801& 18.86/0.389& 25.51/0.812& 25.62/0.818& 26.31/0.801 \\
      \cline{2-9}
            & Smiling-crowd& 21.81/0.493& 20.56/0.428& 25.93/0.836& 18.84/0.359& 26.11/0.853& 25.97/0.851& 29.27/0.832 \\
      \cline{2-9}
            & Electro-devices& 22.67/0.322& 21.17/0.260& 28.41/0.824& 19.28/0.201& 28.67/0.836& 28.26/0.823& 30.74/0.793 \\
      \clineB{2-9}{2.0}
            & mean
                     & 22.29/0.415& 20.91/0.352& 27.82/0.842& 19.13/0.288& 28.04/\textbf{0.854}& 27.88/0.852& \textbf{29.66}/0.814\\
      \hline\hline
      \multirow{6}{*}{20/5}& Dishes& 18.89/0.283& 17.51/0.234& 26.15/0.671& 15.31/0.177& 25.31/0.629& 26.18/0.663& 30.36/0.843 \\
      \cline{2-9}
            & Rooster-clock& 19.19/0.267& 17.69/0.212& 27.91/0.710& 15.42/0.158& 27.18/0.680& 27.80/0.720& 31.39/0.794 \\
      \cline{2-9}
            & Coffee-beans-vases& 18.25/0.394& 16.99/0.342& 23.56/0.646& 14.96/0.274& 23.13/0.625& 23.81/0.652& 26.22/0.797 \\
      \cline{2-9}
            & Smiling-crowd& 18.14/0.383& 16.86/0.329& 23.97/0.693& 14.53/0.249& 23.17/0.658& 23.86/0.695& 29.12/0.829 \\
      \cline{2-9}
            & Electro-devices& 18.93/0.231& 17.51/0.184& 26.00/0.631& 15.30/0.135& 25.29/0.595& 25.76/0.621& 30.65/0.789 \\
      \clineB{2-9}{2.0}
            & mean
                     & 18.68/0.312& 17.31/0.260& 25.52/0.670& 15.10/0.199& 24.82/0.638& 25.48/0.670& \textbf{29.55}/\textbf{0.810}\\
      \hline\hline
      \multirow{6}{*}{50/0}& Dishes& 16.11/0.162& 14.56/0.129& 20.81/0.819& 11.81/0.086& 20.51/0.818& 20.62/0.821& 27.74/0.715 \\
      \cline{2-9}
            & Rooster-clock& 15.90/0.133& 14.28/0.101& 21.62/0.787& 11.43/0.065& 21.36/0.789& 21.28/0.793& 28.55/0.669 \\
      \cline{2-9}
            & Coffee-beans-vases& 15.92/0.246& 14.44/0.196& 19.85/0.693& 11.83/0.138& 19.69/0.694& 19.71/0.709& 24.26/0.672 \\
      \cline{2-9}
            & Smiling-crowd& 16.28/0.247& 14.80/0.203& 19.02/0.686& 12.13/0.146& 18.88/0.695& 18.91/0.692& 26.40/0.717 \\
      \cline{2-9}
            & Electro-devices& 15.99/0.120& 14.41/0.093& 21.10/0.743& 11.60/0.062& 20.92/0.745& 20.81/0.732& 28.20/0.662 \\
      \clineB{2-9}{2.0}
            & mean
                     & 16.04/0.182& 14.50/0.145& 20.48/0.746& 11.76/0.100& 20.27/0.748& 20.27/\textbf{0.749}& \textbf{27.03}/0.687\\
      \hline\hline
      \multirow{6}{*}{50/20}& Dishes& 13.00/0.094& 11.78/0.075& 18.87/0.743& 9.50/0.047& 18.66/0.736& 18.85/0.745& 26.89/0.704 \\
      \cline{2-9}
            & Rooster-clock& 13.23/0.080& 11.91/0.061& 19.76/0.721& 9.48/0.038& 19.56/0.714& 19.40/0.721& 27.80/0.675 \\
      \cline{2-9}
            & Coffee-beans-vases& 12.78/0.145& 11.61/0.115& 17.82/0.622& 9.45/0.077& 17.69/0.618& 17.60/0.635& 23.78/0.674 \\
      \cline{2-9}
            & Smiling-crowd& 12.66/0.148& 11.52/0.121& 16.44/0.621& 9.44/0.083& 16.33/0.622& 16.39/0.625& 25.54/0.721 \\
      \cline{2-9}
            & Electro-devices& 13.05/0.068& 11.78/0.053& 19.40/0.690& 9.45/0.033& 19.25/0.684& 19.09/0.677& 27.44/0.656 \\
      \clineB{2-9}{2.0}
            & mean
                     & 12.94/0.107& 11.72/0.085& 18.46/0.680& 9.46/0.056& 18.30/0.675& 18.26/0.681& \textbf{26.29}/\textbf{0.686}\\
      \hline
    \end{tabular}
    \label{tab:sta}
  \end{threeparttable}
\end{table*}
The experimental results are reported in Table~\ref{tab:sta} and visualized in Fig.~\ref{fig:sr_x2_sta}. For the two approaches, resLF and 3DVSR, which do not support noisy LF input, we generate de-noised LF with DRLF and use it as an input to resLF and 3DVSR. These results are denoted as \textit{De+resLF} and \textit{De+3DVSR} respectively.
For all noise settings, our approach provides the best reconstruction quality in terms of PSNR. For mixed noise settings, the proposed method achieves an averagely highest SSIM score as compared to the other approaches. These high scores pay tribute to the robustness of the proposed model in which de-noising and super-resolution are jointly resolved.
Without de-nosing resLF and 3DVSR completely fails to reconstruct a good quality HR image. In practice, they up-scale not only the texture but also the existing noise. Their scores are, therefore, even worse as compared to bi-cubic upsampling approach in which noise are blurred out.
From Fig.~\ref{fig:sr_x2_sta}, it is evident that the reconstructed HR image from the other approaches is over-smoothed while our approach preserves well the texture content and high-frequency information, e.g.,  and object edges in \textit{Dishes} scene, background pattern in \textit{Smiling-crowd} scene.
Since DRLF supports only Gaussian noise, it fails to recognize impulse noise in the LR input. The impulse noise is either ignored, i.e., when Gaussian noise level is low, or mistreated, i.e., in a severe Gaussian noise setting.
Consequently, the reconstructed HR images are presented with noisy traces, i.e., \textit{Coffee-beans-vases} scene or losing texture detail, i.e., the flower bud in \textit{Dishes} scene.
\subsection{Comparison to GPU-Accelerated Approach}
\begin{figure}[t]
  \centering
  \begin{minipage}[b]{.9\linewidth}
    \centering
    \includegraphics[width=\textwidth]{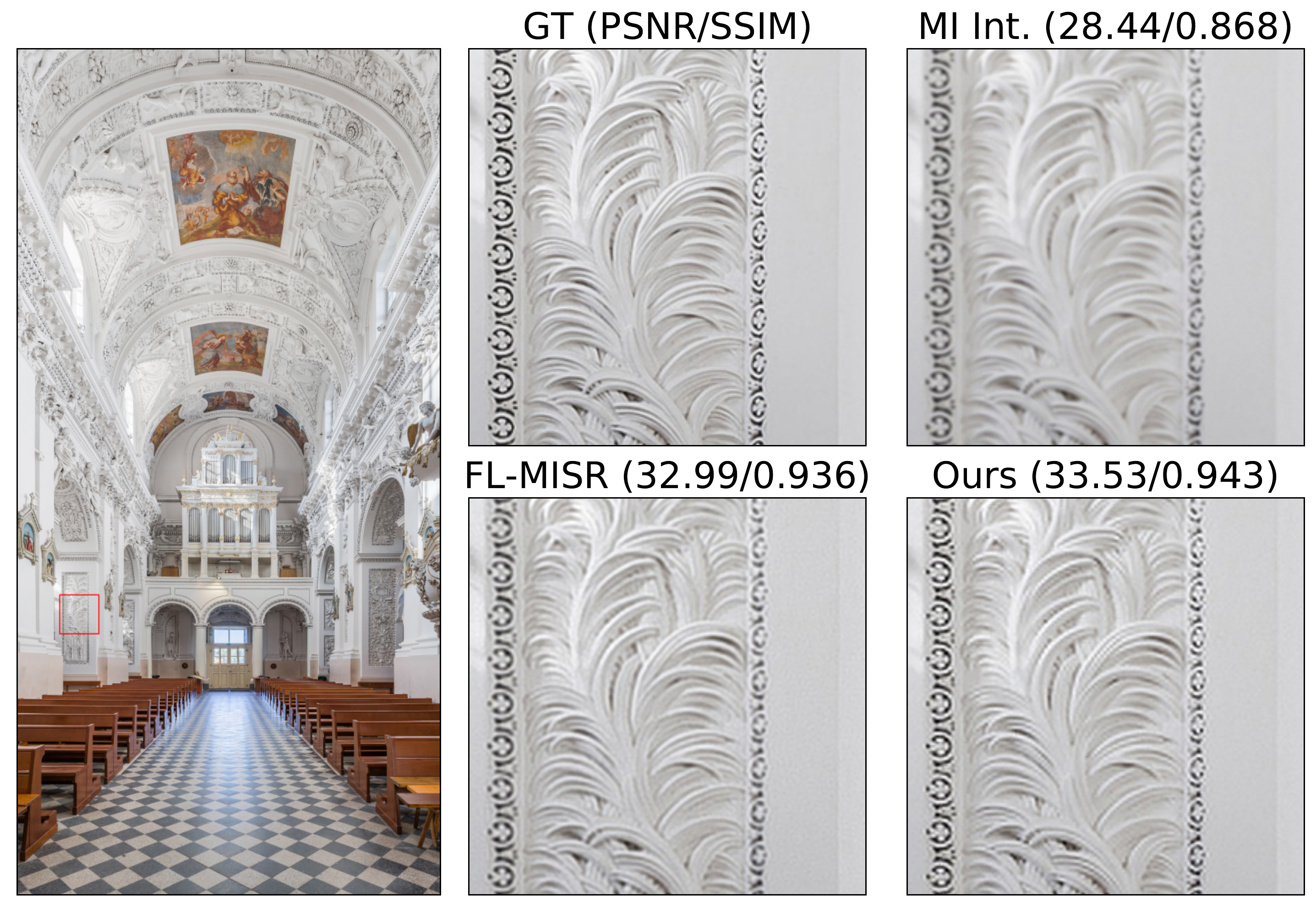}
  \end{minipage}
  \begin{minipage}[b]{.9\linewidth}
    \centering
    \includegraphics[width=\textwidth]{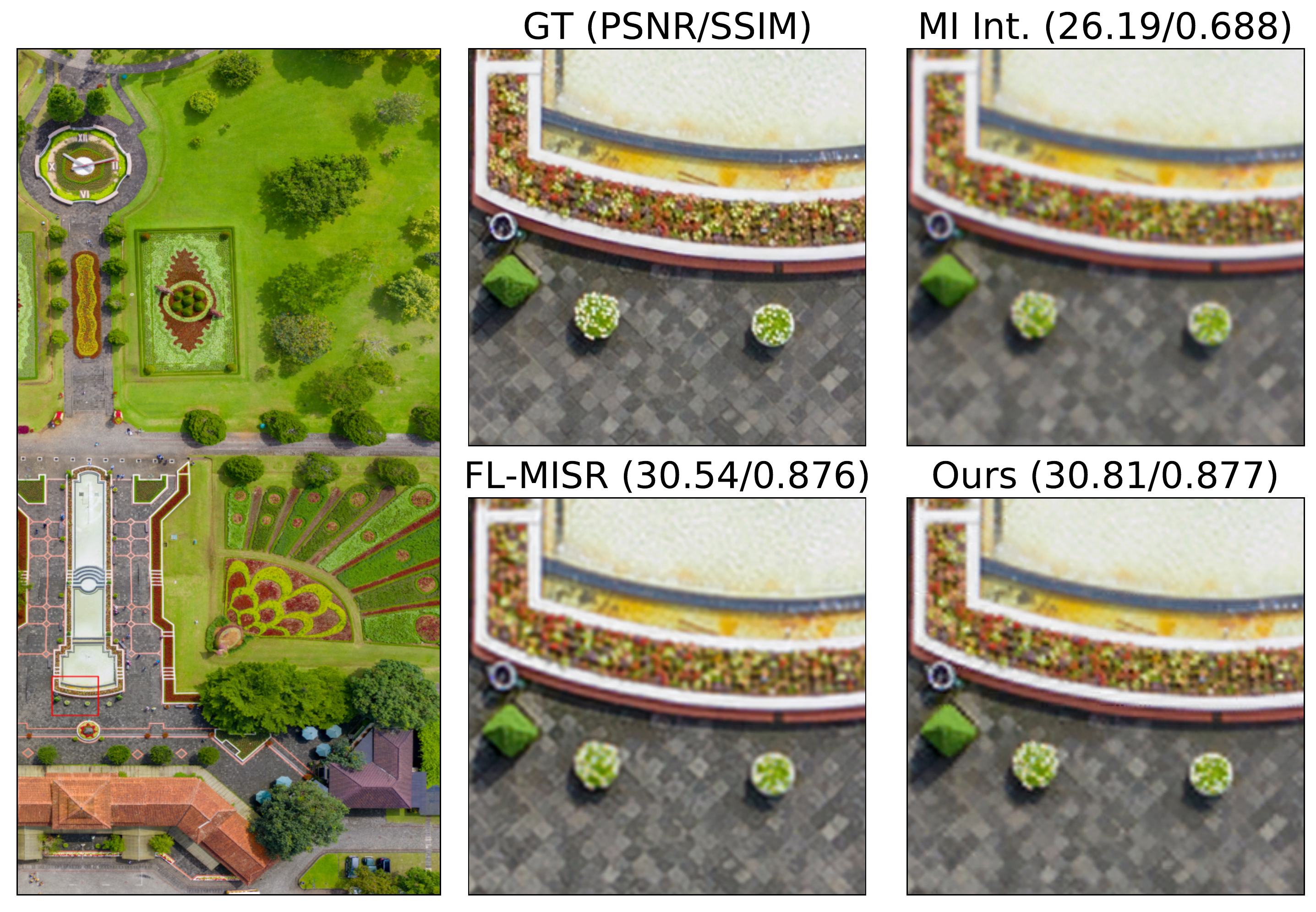}
  \end{minipage}
  \caption{HR reconstruction results of DIV8K
    dataset~\cite{Gu2019div8k}. \textit{top} $\Stimes 2$ results of
    image 0002; \textit{bottom}  $\Stimes 3$ results of image 0084.}
  \label{fig:mfsr}
\end{figure}

As discussed in Sec.~\ref{sec:sr_degrad}, the proposed framework shares a similar setup as a multi-frame super-resolution problem and indeed can be applied as well for this kind of problem.
To evaluate the performance of our accelerated framework, we conducted an experiment on the natural image dataset DIV8K~\cite{Gu2019div8k} and compared to recent related work on the field (FL-MISR~\cite{Sun2021fl}).
We follow the experimental setup described in~\cite{Sun2021fl} to prepare the low-resolution images and perform the HR image reconstruction with our accelerated solver.
Particularly, we pick up seven images from DIV8K dataset and generate, for each of them, four LR images for $\Stimes 2$ SR and nine LR images for $\Stimes 3$ SR.
The shifting of $\Stimes 2$ and $\Stimes 3$ image sets are respectively $\frac{1}{2}$px and $\frac{1}{3}$px.
The Gaussian noise is configured with $\sigma=1$.
Since FL-MISR use $\ell^1$ data fidelity and BTV regularization in their model, we turn off our $\ell^2$ term and configure nonlocal weighting (i.e. $W_\bd$) to match BTV condition.
The accelerated ADMM iterative solver is then executed to minimize the cost function in Eq.~\ref{eq:sr_fin}. For a fair comparison, we stop our iterative solver as soon as the quality of the reconstructed image is comparable to FL-MISR and measure the execution time. Quantitative evaluation results are listed in
Table~\ref{tab:mfsr}, while visual comparison is given in Fig.~\ref{fig:mfsr}
From the table, it is obvious that our GPU accelerated solver outperforms FL-MISR in processing speed for all test cases while providing a better reconstruction quality.
As compared to FL-MISR, our GPU-based solver achieves an average speed-up of 2.46$\Stimes$ and 1.57$\Stimes$ for up-scaling $\Stimes 2$ and up-scaling $\Stimes 3$ respectively.
This performance boost tributes to the effectiveness of ADMM solver and the realization strategy of transformation matrices ($A_k, S_\bd$).
In contrast to FL-MISR, which chooses to implement $A_k, S_\bd$ with sparse matrices, our approach takes advantage of linear functions (i.e., $\opW, \opB, \opD$) to optimize GPU memory and computation resource.
Therefore, our GPU-based solver can fit well within a single GTX 1080Ti GPU, while FL-MISR needs four of them to solve the same problem.

\begin{table*}[!h]
  \centering
  \setlength{\tabcolsep}{2pt}
  \captionsetup{justification=centering}
  \begin{threeparttable}
    \vspace{8pt}
    \caption{Evaluation of parallel computing approach for MISR problem on 8-bit
      natural images in DIV8K dataset. MI Int.:
      Multi-image interpolation (56 cores Intel Xeon Gold 5120), FL-MISR~\cite{Sun2021fl} (4 GTX
      1080Ti), ours (1 GTX 1080Ti).
    }
    \begin{tabular}{ | c | c | c | c | c | c | c | c | c | }
      \hline
      \multicolumn{2}{ |c|}{Image Index} & 0001 &0002 & 0007 & 0027 & 0055 & 0066 & 0084 \\ \hline
      \multicolumn{2}{ |c|}{Resolution of GT} &  5376$\times$5760 & 5568$\times$5760 & 1920$\times$2880 & 2112$\times$2880 & 5760$\times$5760 & 1920$\times$2880 & 5760$\times$3840\\  \hline
      \hline
      \multicolumn{9}{ |c| }{Upscaling 2$\times$} \\  \hline
      \hline
      \multirow{2}{*}{MI Int.} &PSNR/SSIM &30.49/0.9215 &28.44/0.8677 &33.68/0.8810 &28.37/0.8988 &33.80/0.9018 & 35.21/0.9296 &29.11/0.8277\\
                                         & Runtime ($s$) & 0.51&0.52 &0.11 &0.20 &0.53 &0.11 &0.36\\ \hline
      \multirow{2}{*}{FL-MISR} &PSNR/SSIM &37.11/0.9620 &32.99/0.9360 &35.09/0.9111 &33.21/0.9417 &38.03/0.9564 & 37.12/0.9452 &34.13/0.9410\\
                                         & Runtime ($s$) &1.50 &1.29 &0.69 &0.71 &1.3 &0.66&1.21\\ \hline
      \multirow{2}{*}{Ours} &PSNR/SSIM &37.24/0.9713 &33.53/0.9430 &35.42/0.9220 &33.73/0.9497 &38.13/0.9616 &37.51/0.9539 &34.60/0.9454\\
                                         & Runtime ($s$) &0.92 &1.14 &0.15 &0.24 &0.72 &0.18 &0.83\\ \hline
      \hline
      \multicolumn{9}{ |c| }{Upscaling 3$\times$} \\  \hline
      \hline
      \multirow{2}{*}{MI Int.} &PSNR/SSIM &26.74/0.8460 &25.65/0.7749 &32.03/0.8395 &25.15/0.8212 &30.79/0.8153 &32.65/0.8968 &26.19/0.6883 \\
                                         & Runtime ($s$)&1.00 &0.99 &0.11 &0.13 &0.55 & 0.11&0.38\\ \hline
      \multirow{2}{*}{FL-MISR} &PSNR/SSIM &33.24/0.9446 &29.43/0.8941 &33.99/0.8941 &30.17/0.9139 &35.90/0.9379 &36.06/0.9398 &30.54/0.8764 \\
                                         & Runtime ($s$)&1.78 &1.73 &0.32 &0.38 &1.93 &0.35&1.65\\ \hline
      \multirow{2}{*}{Ours} &PSNR/SSIM &33.39/0.9517 &30.04/0.9017 &34.50/0.8984 &30.57/0.9293 &36.22/0.9402 &36.35/0.9447 &30.81/0.8774\\
                                         & Runtime ($s$)&1.13 &1.17 &0.23 &0.25 &1.22 &0.23 &0.84\\ \hline
    \end{tabular}
    \vspace{-10pt}
    \label{tab:mfsr}
  \end{threeparttable}
\end{table*}
\subsection{Performance Analysis of OpenCL-based Solvers}
\begin{figure}
  \centering
  \includegraphics[width=0.99\linewidth]{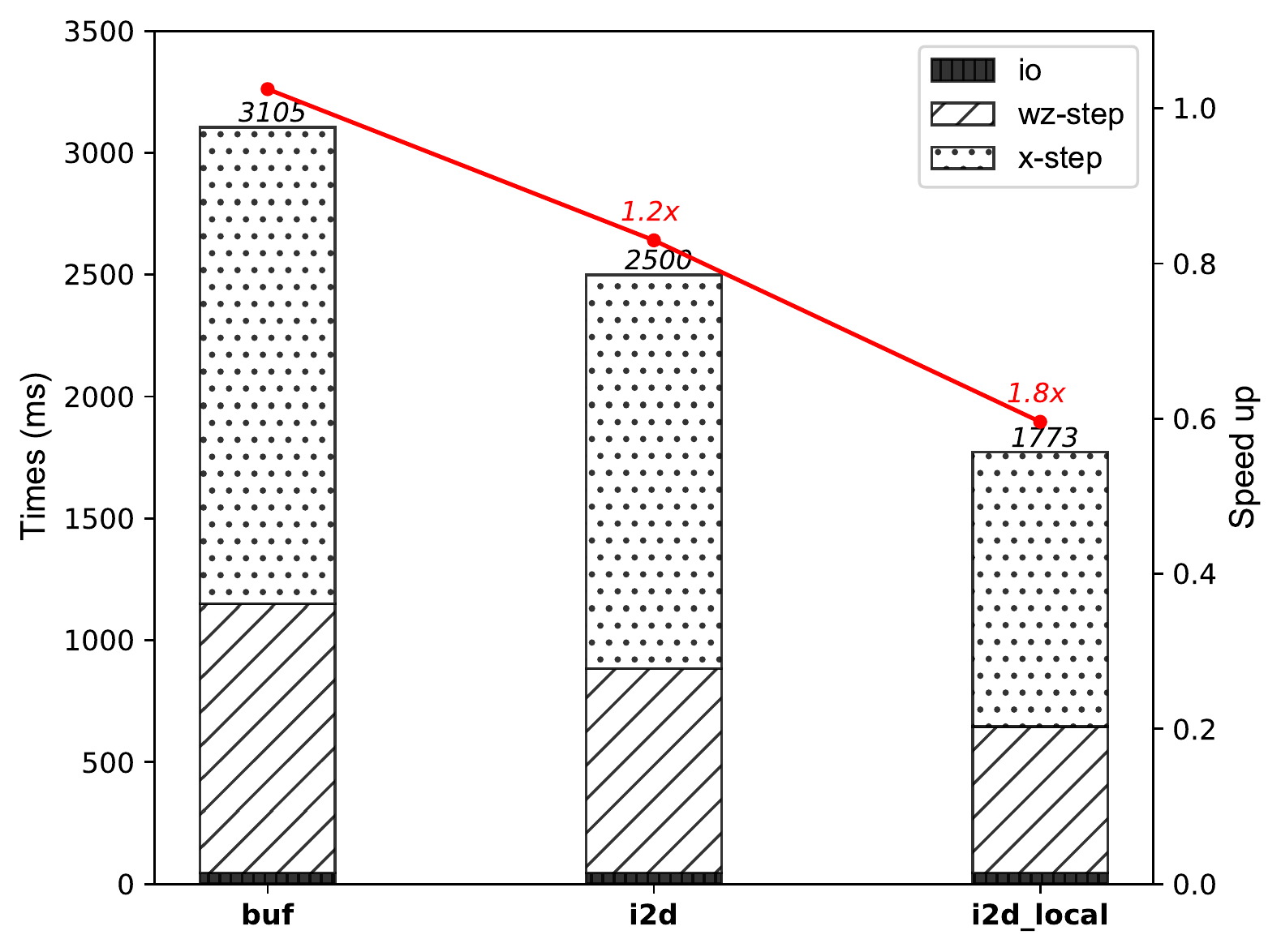}
  \captionsetup{belowskip=-15pt}
  \caption[]{Cumulative execution time and speed-up of three realization strategies.}
  \label{fig:sr_admm_buf_i2d_local}
\end{figure}
\begin{figure}
  \centering
  \includegraphics[width=0.99\linewidth]{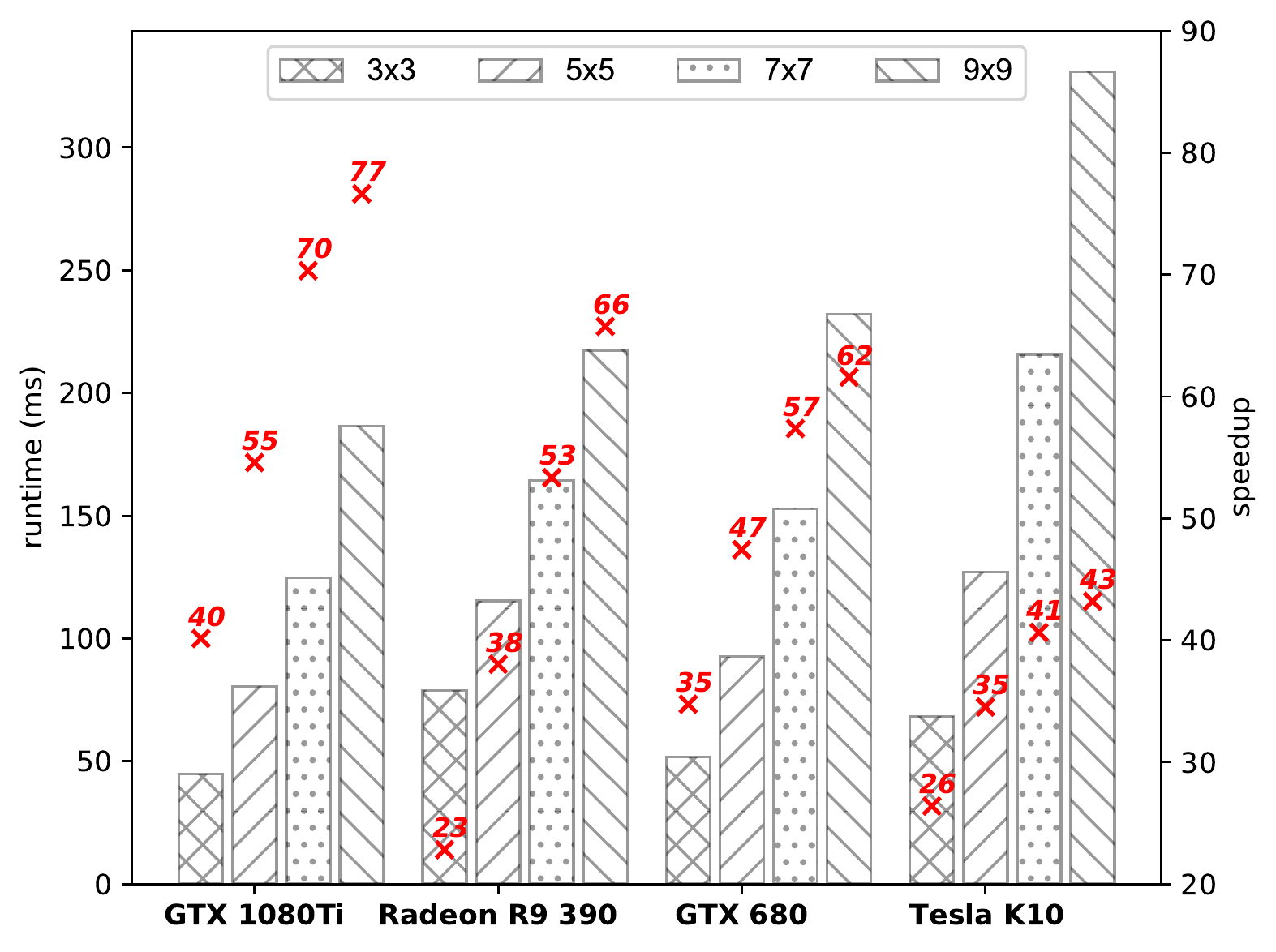}
  \captionsetup{belowskip=-18pt}
  \caption[]{Execution of GPU-based solver under different number of
    inputs on various OpenCL platforms.}
  \label{fig:sr_admm_platforms}
\end{figure}
For analyzing the performance improvement of the proposed GPU accelerated approach, we perform an evaluation of three realization strategies of ADMM iterative solvers. Fig.~\ref{fig:sr_admm_buf_i2d_local} reports the cumulative execution time of the three GPU implementations.
The initial GPU implementation (i.e., \texttt{buf}) is considered as a baseline, in which a 1D buffer object is used for holding variable and input data in GPU global memory.
In the second implementation, denoted as $i2d$, 1D buffer objects are replaced by Image2D objects. This allows us to make use of the texture cache provided in GPU architecture for speeding up the access to image-like data.
The third implementation, denoted as $i2d\_local$ takes advantage of local memory for buffering and sharing data within a work-group.
Since local memory is close to the computing unit, this provides a high-speed data pool for kernel tasks which frequently require access to multiple neighbor pixels (i.e. blurring, warping).
For this experiment, we use $5\Stimes 5$ angular views as input for $\Stimes 4$ SR to a spatial resolution of $512\Stimes 512$. The number of ADMM iterations and CG iterations is set to 10 and 5, respectively.
The execution time of the ADMM solver can be divided into three parts. The $io$ part covers the time for transferring input data from CPU memory into GPU global memory and reading back the reconstructed HR image from GPU to GPU memory.
The \textit{$wz-step$} part represents the computation time of updating $\bw$ and $\bz$ in an ADMM iteration, while \textit{$x-step$} part measures the time to solve for $\xx$ by applying conjugate gradient descent technique, see Fig.~\ref{fig:sr_gpu_admm}.
As could be seen from Fig.~\ref{fig:sr_admm_buf_i2d_local}, the IO time only accounts for a small amount of overall execution time, while most of the time is spent on \textit{$x-step$} and \textit{$wz-step$}.
As compared to the $buf$ version, the texture cache provided by Image2D object $i2d$ does shorten the computation time of ADMM solver by a factor of 1.2$\times$.
The local memory sharing technique further speeds up the computation time by a factor of 1.8$\times$.

The advantage of using the OpenCL framework is that the accelerated solver can be executed on various platforms. Fig.~\ref{fig:sr_admm_platforms} shows the execution time of $i2d\_local$ on various GPU platforms.
In this test, we vary the number of input LR images: 9, 25, 49, and 81, which
are denoted as $3\times3$,$5\times5$,$7\times7$, and $9\times9$, respectively.
The regularization window size is configured to $5\Stimes 5$, and the number of conjugate gradient steps is set to 5.
For each case, we measure the execution time of a single ADMM iteration and
compare it to the CPU implementation, executed on i7-5820K 3.30GHz.
In general, we observed a higher speed up as compared to CPU execution when more
input images are provided. The speedup ranges from 23$\times$ to 40$\times$ in
the case of $3\times 3$ inputs and from 43$\times$ to 77$\times$ in the case of
$9\times 9$ inputs.
\section{Conclusion}
This paper presents a GPU-accelerated computational framework for reconstructing high-resolution SAI from 4D LF data under mixed Gaussian-Impulse noise conditions.
The proposed SR model derived from a statistical perspective takes advantage of a joint $\ell^1-\ell^2$ data fidelity term for dealing with mixed noise conditions and weighted non-local total variation for enforcing LF image prior. Our approach combines the de-noising effect and SR reconstruction into a single optimization problem which, as shown in the experimental results, allows us to surpass the current state-of-the-art approaches in which de-noise and SR problems are resolved separately.
The non-smooth convex optimization problem resulting from the proposed SR model is effectively solved by ADMM algorithm. By transforming the minimization of $\ell^1 -\ell^2-\ell^1$ mixture cost function into least square approximation and proximal operator problems, ADMM overcomes the main problem of gradient descent technique in finding a suitable step-size. We showed that GPU acceleration is well-suited to speeding up the iteratively
solving process.
To verify the robustness of the proposed SR model and evaluate the performance of the accelerated optimizer, an extensive experiment is conducted on 4D synthetic LF dataset and high-resolution natural image dataset. The experimental results show that the proposed approach outperforms the previous work in accelerating the super-resolution task and optimizing GPU resources. While providing a better reconstruction quality, our accelerated framework provides an average speed up of 2.46$\Stimes$ and $1.57\Stimes$ for $\Stimes 2$ and $\Stimes 3$ SR tasks, respectively.
The accelerated solver achieves a speedup of 77$\Stimes$ as compared to CPU implementation.

The proposed approach encourages further research directions on both algorithmic and computing architecture levels. In the first direction, we would extend the SR model to handle a more challenging noise setting, i.e., photon noise, which follows the Poisson distribution. Solving such a problem would require a new ADMM decomposing strategy for the non-convex non-smooth optimization problem. In the second direction, the iterative solving process could be realized on a field-programmable gate array (FPGA) platform on which we could achieve much higher processing speed and much lower energy consumption as compared to GPU. For this task, the main challenges lie in the realization of the warping function and the access of 4D-LF data on hardware.

%
%



\bibliographystyle{elsarticle-num}
\bibliography{lfsr}
\end{document}